\PassOptionsToPackage{table}{xcolor}
 \documentclass[sigconf]{acmart}

\usepackage{soul}

\usepackage{enumitem}

\newenvironment{keyfinding}[1]
  {\par\medskip\noindent\textbf{#1.}\itshape}
  {\par\medskip}

\copyrightyear{2026}
\acmYear{2026}
\setcopyright{cc}
\setcctype{by}
\acmConference[MSR '26]{23rd International Conference on Mining Software Repositories}{April 13--14, 2026}{Rio de Janeiro, Brazil}
\acmBooktitle{23rd International Conference on Mining Software Repositories (MSR '26), April 13--14, 2026, Rio de Janeiro, Brazil}
\acmPrice{}
\acmDOI{10.1145/3793302.3793360}
\acmISBN{979-8-4007-2474-9/2026/04}

\title{Characterizing and Modeling  the GitHub   Security Advisories Review Pipeline}

\author{Claudio Segal}
\orcid{0009-0000-4101-4191}
\affiliation{%
  \institution{Institute of Computing,  Fluminense Federal University (IC-UFF)}
  \city{Niter\'oi}
  \state{RJ}
  \country{Brazil}
}

\author{Paulo Segal}
\orcid{0009-0000-3265-3047}
\affiliation{%
  \institution{Institute of Computing, Fluminense Federal University (IC-UFF)}
  \city{Niter\'oi}
  \state{RJ}
  \country{Brazil}
}

\author{Carlos Eduardo Banjar}
\orcid{0000-0003-1576-3863}
\affiliation{%
  \institution{Institute of Computing,  Federal University of Rio de Janeiro (IC-UFRJ)}
  \city{Rio de Janeiro}
  \state{RJ}
  \country{Brazil}
}

\author{Felipe de Sant'Anna Paix\~ao}
\orcid{0009-0009-7370-5478}
\affiliation{%
  \institution{Institute of Computing, Federal University of Bahia (IC-UFBA)}
  \city{Salvador}
  \state{BA}
  \country{Brazil}
}

\author{Hudson Silva Borges}
\orcid{0000-0002-8325-8773}
\affiliation{%
  \institution{Federal University of Mato Grosso do Sul (UFMS)}
  \city{Campo Grande}
  \state{MS}
  \country{Brazil}
}

\author{Paulo Silveira}
\orcid{0000-0003-0197-8249}
\affiliation{%
  \institution{Rural Federal University of  Pernambuco (UFRPE)}
  \city{Recife}
  \state{PE}
  \country{Brazil}
}

\author{Eduardo Santana de~Almeida}
\orcid{0000-0002-9312-6715}
\affiliation{%
  \institution{Institute of Computing, Federal University of Bahia (IC-UFBA)}
  \city{Salvador}
  \state{BA}
  \country{Brazil}
}

\author{Joanna C.\ S.\ Santos}
\orcid{0000-0001-8743-2516}
\affiliation{%
  \institution{University of Notre Dame}
  \city{Notre Dame}
  \state{IN}
  \country{USA}
}

\author{Anton Kocheturov}
\orcid{0000-0003-2549-9146}
\affiliation{%
  \institution{Siemens Corporation}
  \city{Princeton}
  \state{NJ}
  \country{USA}
}

\author{Gaurav Srivastava}
\orcid{0009-0005-2755-629X}
\affiliation{%
  \institution{Siemens Corporation}
  \city{Princeton}
  \state{NJ}
  \country{USA}
}

\author{Daniel S.~Menasch\'e}
\orcid{0000-0002-8953-4003}
\affiliation{%
  \institution{IC-UFRJ}
  \city{Rio de Janeiro}
  \state{RJ}
  \country{Brazil}
}

\AtBeginDocument{%
  \markboth{\shorttitle}{Segal et al.}
}

\usepackage{makecell}
\newcolumntype{Y}{>{\raggedright\arraybackslash}X}

\setlist[itemize]{noitemsep, topsep=0pt, leftmargin=*}

\copyrightyear{2026}
\acmYear{2026}
\setcopyright{cc}
\setcctype{by}
\acmConference[MSR '26]{23rd International Conference on Mining Software Repositories}{April 13--14, 2026}{Rio de Janeiro, Brazil}
\acmBooktitle{23rd International Conference on Mining Software Repositories (MSR '26), April 13--14, 2026, Rio de Janeiro, Brazil}
\acmPrice{}
\acmDOI{10.1145/3793302.3793360}
\acmISBN{979-8-4007-2474-9/2026/04}

\begin{document}
\begin{abstract}
\emph{GitHub Security Advisories} (GHSA) have become a central component of open-source vulnerability disclosure and are widely used by developers and security tools. A distinctive feature of GHSA is that only a fraction of advisories are \emph{reviewed} by GitHub, while the mechanisms associated with this review process remain poorly understood. In this paper, we conduct a large-scale empirical study of the GHSA review processes, analyzing over 288,000 advisories spanning 2019--2025. We characterize which advisories are more likely to be reviewed, quantify review delays, and identify two distinct review-latency regimes: a \emph{fast path} dominated by GitHub Repository Advisories (GRAs) and a \emph{slow path} dominated by NVD-first advisories. We further develop a queueing model that accounts for this dichotomy based on the structure of the advisory processing pipeline.
\end{abstract}


\maketitle

\markboth{Segal et al.}{Segal et al.} 

\section{Introduction}
The security of modern software ecosystems depends on the timely identification, disclosure, and remediation of vulnerabilities in open-source components \cite{ayala2025investigating, ayala2024poster, ayala2025mixed, ayala2025deep}. The GitHub Security Advisories (GHSA)~\cite{AboutGithubAdvisoryDatabase} database plays a central role in this process by providing structured vulnerability information and integrating with tools such as Dependabot~\cite{AboutDependabotAlets,he2023automating}, which automatically alerts developers when dependencies are affected by known vulnerabilities. Given the scale of open-source adoption, the GHSA database is directly relevant to millions of users worldwide~\cite{GitHubAdvisoryDatabase,GHSADByTheNumbers}.

Within GHSA, some advisories undergo a formal \emph{GitHub review}, which validates the report, enriches it with metadata, and increases its visibility in automated pipelines, while others remain unreviewed. Reviews can substantially influence how quickly and effectively vulnerabilities are triaged and patched \cite{AboutDependabotAlets, he2023automating}.

While previous studies have examined various facets of the security advisory publishing process, including disclosure practices~\cite{arora2010empirical}, automation~\cite{he2023automating}, and governance~\cite{ayala2025mixed,alexopoulos2021vulnerability},  little is known about the factors that drive the advisory reviews. How long do reviews take once advisories are published? And do review patterns reveal implicit prioritization strategies?

This paper addresses these questions through a systematic analysis of \textbf{288,604} GHSA vulnerability reports, including \textbf{23,563} that received GitHub reviews as of August 21, 2025. We quantify the timing of reviews, propose models for predicting review outcomes, and discuss implications for vulnerability management. Understanding review timing has both academic and industry value: it enables organizations to align their triage strategies with ecosystem dynamics and to anticipate which vulnerabilities are more likely to receive timely attention \cite{ayala2025investigating, AboutDependabotAlets}.
 
The contributions of this work are as follows:
\begin{itemize}[leftmargin=*, topsep=0pt, partopsep=0pt]
\item We analyze the roles and maturity of users credited in GitHub-reviewed advisories, and study how these roles correlate with advisory review processes (Section~\ref{sec:usersroles}), offering insights into the human factors that shape them.
\item We provide the first large-scale characterization of GHSA reviews, examining their frequency, distribution across ecosystems and severities, and time-to-review dynamics (Section~\ref{sec:empirical}).
\item We identify distinct \emph{fast} and \emph{slow} review paths in the advisory pipeline (Sections~\ref{sec:empirical}--\ref{sec:model}), showing that advisories originating as GRAs tend to be reviewed earlier than NVD-first advisories. However, our information flow analysis (Section~\ref{sec:usersroles} and Figure~\ref{fig:sankey}) reveals substantial room for improving how GRAs are currently leveraged, as most GHSAs still do not originate from GRAs.
\item We develop a simple analytical model of advisory review timing (Section~\ref{sec:model}),
capturing the essence of the \emph{fast} and \emph{slow}   paths observed in the GHSA pipeline, and enabling what-if analyses.
\item We discuss how these insights can inform vulnerability triage, policy considerations, and automated tooling (Section~\ref{sec:discussion}).
\item We release a curated dataset and scripts to enable replication and future research: \url{https://github.com/cmsegal/ghsa-review}.
 \end{itemize} 

\textbf{Outline.}
 The remainder of this paper is organized as follows.
Section~\ref{sec:background} gives a background. Section~\ref{sec:miningmethodology} describes our mining methodology and dataset construction.
Section~\ref{sec:usersroles} examines the roles, popularity, and engagement patterns of users involved in security advisories, Section~\ref{sec:empirical} provides an empirical characterization of the review pipeline, and  
Section~\ref{sec:model} develops a queueing-based model that captures key statistical properties of the review process, 
Section~\ref{sec:discussion} discusses the practical implications of our findings,  Section~\ref{sec:relatedwork} situates our contributions within the broader literature, Section~\ref{sec:threatstovalidity} lists threats to validity and Section~\ref{sec:conclusion} concludes this paper. 

\section{Background}\label{sec:background}

The \textbf{\textit{GitHub Security Advisory Database (GHSA)}} is a comprehensive repository of known security vulnerabilities in open source software packages. The database categorizes advisories into two main groups: \textit{GitHub-reviewed advisories} and \textit{unreviewed advisories}~\cite{GHSADByTheNumbers}. GitHub-reviewed advisories are security vulnerabilities that have been manually validated and mapped to packages in supported ecosystems, ensuring they contain complete descriptions and accurate ecosystem information. Unreviewed advisories  remain in this category either because they are pending review, do not affect supported packages, or do not describe valid vulnerabilities \cite{GitHubAdvisoryDatabase}.

Reviewed advisories originate from multiple sources, with the two most significant being the \textit{National Vulnerability Database (NVD)}~\cite{NVD} and the \textit{GitHub Repository Advisories (GRA) }\cite{GHSADByTheNumbers}. The NVD is the largest source, covering vulnerabilities across all software types, while GitHub reviews only those relevant to its supported ecosystems. GRAs constitute the second-largest source and provide maintainers with a collaborative workflow for responsible vulnerability disclosure \cite{GithubBlogAdvisoryCredits}. Maintainers can create draft advisories to privately discuss security issues, develop fixes in temporary private forks, and publish them to alert their communities. Once published, GitHub reviews each advisory and incorporates it into the GHSA database. These advisories may trigger Dependabot alerts, notifying repository owners when dependencies contain known vulnerabilities~\cite{AboutDependabotAlets}.

\section{Mining Methodology and Goals}
\label{sec:miningmethodology}

Our study is based on the GHSA dataset as of August 21, 2025, which provides structured metadata for each advisory, including its identifier (\texttt{ghsa\_id}), affected repositories (\texttt{source\_code\_location}), and a Common Vulnerabilities and Exposures (CVE) identifier (\texttt{cve\_id}), which is a unique identifier within NVD. Using GitHub’s global advisories API~\cite{GithubGlobalAdvisoriesAPI}, we extracted \textbf{288,604} advisories, of which \textbf{23,563} are GitHub-reviewed and the rest unreviewed.

Temporal fields are central to our study: \texttt{published\_at} marks the advisory's first appearance in GHSA, \texttt{nvd\_published\_at} records the corresponding date in NVD, and \texttt{github\_reviewed\_at} indicates when GitHub completed its internal review process. Additional fields capture affected product ranges (\texttt{vulnerabilities}), supporting references (\texttt{references}), and user credits (\texttt{credits}), later analyzed to understand contributor roles.

\begin{table}[!htbp]
\caption{GHSA dataset fields, including enrichment.}
\label{tab:fields}
\footnotesize
\begingroup
\setlength{\aboverulesep}{0pt}
\setlength{\belowrulesep}{0pt}
\setlength{\tabcolsep}{2pt} 
\rowcolors{2}{white}{gray!10}
\begin{tabular}{lp{5.4cm}}
\toprule
\textbf{Field} & \textbf{Description} \\
\midrule
\texttt{ghsa\_id} & GHSA identifier. \\
\texttt{source\_code\_location} & Link to reported GitHub repository~\cite{kancharoendee2025categorizing}. \\
\texttt{repository\_advisory\_url} & URL of the originating GRA, if any. \\
\texttt{published\_at} & Publication date of the GHSA. \\
\texttt{nvd\_published\_at} & Publication date in the NVD, if imported. \\
\texttt{github\_reviewed\_at} & GitHub review date (if reviewed). \\
\texttt{cve\_id} & Associated Common Vulnerabilities and Exposures (CVE) identifier. \\
\texttt{severity} & Severity of the advisory (low, medium, high, critical or unknown). \\
\texttt{references} & External references regarding the advisory. \\
\texttt{vulnerabilities} & Products and affected version ranges. \\
\texttt{credits} & List of users and their roles. \\
\midrule
\texttt{gra\_published\_at} & Publication date of the originating GRA (when available). \\
\texttt{enrich\_nvd\_published\_at} & NVD publication date obtained via CVE query \\
& (if missing in GHSA). \\
\texttt{ecosystem\_published\_at} & Dates from ecosystem DBs (RustSec, PyPA, RubySec, etc.). \\
\texttt{patched\_at} & Date of first patched package version. \\
\midrule
\texttt{user\_profile} & Metadata of credited users  (e.g., account age, \#followers, \#public repos). \\
\texttt{repo\_metadata} & Characteristics of affected repositories  (e.g., stars, forks, open issues). \\
\bottomrule
\end{tabular}
\endgroup
\end{table}

\subsection{Data enrichment}  \label{subsec:DataEnrichment}
While the GHSA database provides the temporal metrics aforementioned (e.g., \texttt{published\_at}, \texttt{nvd\_published\_at}, etc.), not all entries in this database may include these metrics. Thus, we conduct an  enrichment process, to fill in key missing values and include additional ones that are not originally present in GHSA.
Table~\ref{tab:fields} summarizes the core fields and the enrichment features derived to trace advisory propagation and patch availability.

Since this study focuses on the GHSA review process, the enrichment is applied only to the \textbf{23{,}563} GitHub-reviewed advisories. This enrichment step aims to provide complementary information on temporal metrics, as well as user and repository metadata.


\subparagraph{\textbf{(a) Temporal enrichment:}} To obtain a more complete and consistent timeline of advisory disclosure events, we enrich  the records with additional publication timestamps from multiple data sources. These timestamps allow measuring propagation across sources and delays between patch release and review.

\begin{enumerate}[label=(\roman*), leftmargin=15pt, noitemsep, topsep=0pt]
    \item If the \texttt{repository\_advisory\_url} field is nonempty, we query the GitHub API for the associated Repository Security Advisory (GRA) and record its publication timestamp as \texttt{gra\_published\_at}.

    \item If the \texttt{nvd\_published\_at} field is missing, we fill it with the NVD-provided publication timestamp, obtained by querying the NVD API using the advisory's \texttt{cve\_id}.

    \item We parse the \texttt{references} field of each advisory to detect links to RustSec, FriendsOfPHP, PyPA, RubySec, and GoVulnDB. For each, we approximate the publication date by retrieving the timestamp of the first commit that introduced the advisory into the corresponding repository. These dates are stored as \texttt{rustsec\_published\_at}, \texttt{pypa\_published\_at}, etc.

    \item For selected ecosystems (pip/PyPI, Go, RubyGems, npm, Maven, NuGet), we extract the first patched package version from the advisory’s \texttt{vulnerabilities} attribute. We then query the deps.dev API to retrieve the \texttt{publishedAt} timestamp of that version, denoted as \texttt{patched\_at}. If multiple packages are affected, we sample the first one listed in the \texttt{vulnerabilities} array (see Section~\ref{sec:time_from_patch_to_review}).
\end{enumerate}





\subparagraph{\textbf{(b) User and repository enrichment:}}
To capture the social and technical context of reviews, we also enrich reviewed advisories with user and repository metadata. For each credited user, we query the GitHub API to retrieve profile metadata, including account creation date, followers, public repositories, and activity indicators, used in Section \ref{sec:usersroles}.  
For repositories linked via \texttt{source\_code\_location}, we collected number of stars, open issues and OpenSSF scores as proxies for project popularity and activity, allowing us to analyze differences in maintenance and security practices.

Together, the user and repository enrichment complement the temporal data, allowing us to view advisory reviews not as isolated events but as part of broader socio-technical ecosystems. The final dataset includes \textbf{5,984} unique GitHub repositories and \textbf{3,928}  users.


\subsection{Research Questions}
Using these data, we address the following research questions:

\begin{itemize}[topsep=0pt, partopsep=0pt,leftmargin=23pt]
    \item[\textbf{RQ1}] \textit{What are the roles of users involved with security advisories?}
    \item[\textbf{RQ2}] \textit{Do users tend to specialize in one role?}
    \item[\textbf{RQ3}] \textit{How does the popularity and engagement   of     actors (e.g., remediation  developers) differ from that of critics (e.g.,  reviewers)?}
    \item[\textbf{RQ4}] \textit{What distinguishes GRA-originated advisories from non-GRA advisories in the characteristics of their linked repositories and reviewers?}
    \item[\textbf{RQ5}] \textit{What are the main sources of GitHub-reviewed advisories, and what is their relative contribution to the review process?}
    \item[\textbf{RQ6}] \textit{What factors influence the time to review an advisory once it is published in GitHub?}
    \item[\textbf{RQ7}] \textit{How long does it take for advisories to be reviewed after the corresponding patches are released?}
\end{itemize}
In the next sections (Sections~\ref{sec:RQ1_RQ4} and ~\ref{sec:RQ5_RQ7}), we present the methodology used to address these RQs along with the corresponding results.

\section{RQ1--RQ4: Users Engagement and Repository Characteristics}
\label{sec:usersroles}\label{sec:RQ1_RQ4}
Understanding how advisories are handled requires examining both the \textit{individuals} and the \textit{projects} behind them. GHSA operates as a socio-technical system, in which user roles and repository characteristics influence the way in which vulnerabilities progress through the review pipeline. Thus, in this section, we analyze user engagement and repository characteristics to explain how social and technical factors influence review outcomes (RQ1--RQ4).


\subsection{User Roles}
\textbf{RQ1:} \emph{What are the roles of users involved with security advisories?}
Security advisories involve multiple user roles, each responsible for specific tasks during the vulnerability lifecycle. By analyzing the \textit{credits} field in GitHub Security Advisories, we identified \textbf{10} distinct roles. Table~\ref{tab:roles} summarizes these roles along with their frequencies and descriptions.


This distribution highlights the collaborative and multifaceted nature of handling security vulnerabilities, with a strong emphasis on validation and reporting roles. These roles align with the standardized contributor categories defined in the CVE 5.0 schema, which formalizes how individuals are credited throughout the vulnerability lifecycle~\cite{GithubBlogAdvisoryCredits,GithubDocsAboutCredits}.


\begin{table}[!htbp]
\centering
\caption{Roles observed in GitHub Security Advisories, with their total frequency of occurrence across advisories and their respective descriptions.}
\label{tab:roles}
\footnotesize
\begingroup
\setlength{\aboverulesep}{0pt}
\setlength{\belowrulesep}{0pt}
\setlength{\tabcolsep}{2pt} 
\rowcolors{2}{white}{gray!10}
\begin{tabular}{p{1.3cm} c p{5.75cm}}
\toprule
\textbf{Role} & \textbf{\# Occurr.} & \textbf{Description} \\
\midrule
Analyst & 5,610 & Validates the vulnerability to confirm its accuracy or severity. \\
Reporter & 2,086 & Notifies the vendor or responsible party about the vulnerability. \\
Finder & 616 & Identifies the vulnerability. \\
Remediation Developer & 602 & Prepares code changes or remediation plans. \\
Remediation Reviewer & 349 & Reviews fixes for effectiveness and completeness. \\
Coordinator & 236 & Facilitates the coordination of the response process. \\
Remediation Verifier & 51 & Tests and verifies that the vulnerability has been properly resolved. \\
Other & 31 & Contributes in ways that do not fit into predefined roles. \\
Sponsor & 7 & Supports the vulnerability identification or remediation process. \\
Tool & 3 & Refers to tools used in the discovery or identification of vulnerabilities. \\
\bottomrule
\end{tabular}
\endgroup
\end{table}

\begin{keyfinding}{Key finding for RQ1}
The vulnerability management process on GitHub is dominated by \textsf{Analysts} and \textsf{Reporters}, while less frequent roles, such as \textsf{Remediation Developers}, \textsf{Reviewers}, and \textsf{Coordinators}, play crucial parts in later remediation stages, highlighting a clear separation between discovery and resolution.
\end{keyfinding}

\subsection{Specialization}
\textbf{RQ2:} \emph{Do users tend to specialize in one role?}
Our analysis of 3,928 users with at least one role revealed that 3,521 (90\%) acted in exactly one role, and 407 (10\%) acted in multiple roles.  Table~\ref{tab:role_distribution} presents the top three most frequent role combinations for users with $n$ distinct roles. Among multi-role users, 77\% combined exactly two roles, with the most frequent combinations being \textsf{Analyst} and \textsf{Reporter} (98 users), \textsf{Finder} and \textsf{Reporter} (47 users), and \textsf{Remediation Reviewer} and \textsf{Remediation Developer} (37 users). Higher role diversity (3--5 roles) typically involved \textsf{Analyst} with remediation-related roles. While most users are specialists, a small group of generalists contribute across multiple roles, potentially bridging tasks across different responsibilities.

\begin{table}[!htbp]
\centering
\caption{Number of users by count of distinct roles, and the top three most frequent role combinations.}
\label{tab:role_distribution}
\footnotesize
\begingroup
\setlength{\aboverulesep}{0pt}
\setlength{\belowrulesep}{0pt}
\setlength{\tabcolsep}{3pt}
\renewcommand{\arraystretch}{1.05}
\rowcolors{2}{gray!10}{white}
\begin{tabular}{c c p{0.73\linewidth}}
\toprule
\textbf{\# Roles} & \textbf{\# Users} & \textbf{Top Three Combinations (Role Sets with Counts)} \\
\midrule
1 & 3521 & -- \\
2 & 313 &
Analyst + Reporter (98)\newline
Finder + Reporter (47)\newline
Remediation Reviewer + Remediation Developer (37) \\
3 & 72 &
Finder + Reporter + Analyst (26)\newline
Remediation Reviewer + Remediation Developer + Analyst (9)\newline
Remediation Developer + Coordinator + Analyst (5) \\
4 & 19 &
Rem.\ Developer + Rem.\ Reviewer + Coordinator + Analyst (4)\newline
Reporter + Coordinator + Analyst + Remediation Reviewer (2)\newline
Remediation Developer + Reporter + Coordinator + Analyst (2) \\
5 & 3 &
Coordinator + Reporter + Rem.\ Reviewer + Rem.\ Developer + Analyst (2)\newline
Coordinator + Reporter + Finder + Rem.\ Developer + Analyst (1) \\
\bottomrule
\end{tabular}
\endgroup
\end{table}


\begin{keyfinding}{Key finding for RQ2}
Most participants in GHSAs are highly specialized, with few (10\%) taking multiple, related roles such as \textsf{Analyst}, \textsf{Reporter}, and \textsf{Remediation}
\end{keyfinding}

\subsection{An Actor-Critic Perspective towards Roles} 

\textbf{RQ3:} \emph{How does the popularity and engagement   of     actors (e.g., remediation  developers) differ from that of critics (e.g.,  reviewers)?} 



Table~\ref{tab:relevance_distribution} presents the distribution of popularity metrics (followers and stars) across contributor roles in security advisories, based on established platform engagement metrics~\cite{cosentino2017,borges2018s}. Statistics are computed at the \textit{credit type occurrence} level, where each user contribution is counted independently. We report median values due to highly skewed distributions with substantial outliers.

\begin{table}[!htbp]
    \centering
    \caption{Popularity and engagement metrics per role}
    \label{tab:relevance_distribution}
    \footnotesize
    \begingroup
    \setlength{\aboverulesep}{0pt}
    \setlength{\belowrulesep}{0pt}

    \rowcolors{3}{white}{gray!10}

    \begin{tabular}{p{1.2cm}ccccccc}
        \toprule
        \textbf{Role} & \textbf{\#}
        & \multicolumn{3}{c|}{\textbf{Stars}} 
        & \multicolumn{3}{c}{\textbf{Followers}} \\
        &  & \textbf{Mean} & \textbf{Med.} & \textbf{Std}
           & \textbf{Mean} & \textbf{Med.} & \textbf{Std} \\
        \midrule
        Analyst & 5610 & 596.16 & 18 & 3521.50 & 178.69 & 36 & 608.10 \\
        Reporter & 2086 & 429.22 & 8 & 2230.90 & 136.42 & 16 & 760.14 \\
        Finder & 616 & 391.76 & 11 & 1825.58 & 176.43 & 22 & 703.94 \\
        \makecell[l]{Remediation\\Developer} & 602 & 1882.82 & 59 & 7386.75 & 724.68 & 93.5 & 2853.51 \\
        \makecell[l]{Remediation\\Reviewer} & 349 & 620.65 & 30 & 1853.61 & 340.77 & 90 & 824.25 \\
        Coordinator & 236 & 1097.08 & 34 & 3532.88 & 785.40 & 122 & 2089.26 \\
        \makecell[l]{Remediation\\Verifier} & 51 & 185.45 & 38 & 336.98 & 222.08 & 132 & 276.53 \\
        Other & 31 & 431.65 & 43 & 973.56 & 208.23 & 58 & 358.67 \\
        Sponsor & 7 & 7050.43 & 0 & 8793.43 & 568.71 & 18 & 701.85 \\
        Tool & 3 & 43.33 & 13 & 64.13 & 52.33 & 15 & 72.60 \\
        \bottomrule
    \end{tabular}
    \endgroup
\end{table}

The credit roles reveal a clear \emph{capability hierarchy} between \emph{actors} (who produce artifacts) and \emph{critics} (who validate them). Discovery-phase actors, \textit{i.e.,} \textbf{finders} and \textbf{reporters}, require minimal platform visibility (median 8--22 followers, under 11 stars), suggesting low barriers to entry. In contrast, \textbf{remediation developers} exhibit markedly higher visibility (median 93.5 followers, 59 stars), reflecting the need for maintainer access and established trust to merge security-critical code.

Among critics, \textbf{analysts} show modest visibility (median 36 followers, 18 stars) similar to discovery actors, while \textbf{remediation reviewers} and \textbf{verifiers} display substantially higher prominence (median 90--132 followers, 30--38 stars), approaching remediation developer levels. This pattern indicates that while vulnerability discovery and initial analysis are broadly accessible, contributing to remediation and review requires greater platform establishment. \textbf{Coordinators} combine high visibility (median 122 followers) with moderate frequency, indicating experienced community members orchestrate the disclosure process.

\begin{keyfinding}{Key finding for RQ3}
Data on the users' roles reveal a capability hierarchy in vulnerability management. Discovery-phase actors (finders, reporters) and early analysts have modest visibility requirements (median 16–36 followers), whereas remediation-phase actors (developers) and later-stage critics (reviewers, verifiers) require substantially greater platform establishment (median 90–132 followers), likely reflecting trust and access controls around security-critical code changes.

\end{keyfinding}

\subsection{Repository Characteristics}

\textbf{RQ4:} \emph{What distinguishes GRA-originated advisories from non-GRA advisories in the characteristics of their linked repositories and reviewers?}

\subsubsection{Dataset} To analyze the repositories referenced by GHSA, we used the \texttt{source\_code\_location} attribute. There are advisories where the same repository URL can appear in different ways; for example, both with the protocol (\texttt{https://}) and without. To address this problem, we normalized the values of this field so that they conform to the format \texttt{github.com/owner/repo}. We also filtered out values that do not refer to a GitHub repository. After this normalization, we ended up with \textbf{5{,}984} distinct GitHub repositories, out of which we analyzed only the \textbf{3{,}779} for which advisories were published after June 2022. For these repositories, using the \texttt{deps.dev} API, we collected the number of stars and open issues, in addition to two security health metrics from the OpenSSF scorecard: the Security-Policy and Maintained scores \cite{OpenSSFScorecard}. Repositories were classified as GRA-linked if they were associated with at least one GRA; otherwise, they were classified as non-GRA-linked.

\begin{figure}
    \centering
    \includegraphics[width=\linewidth]{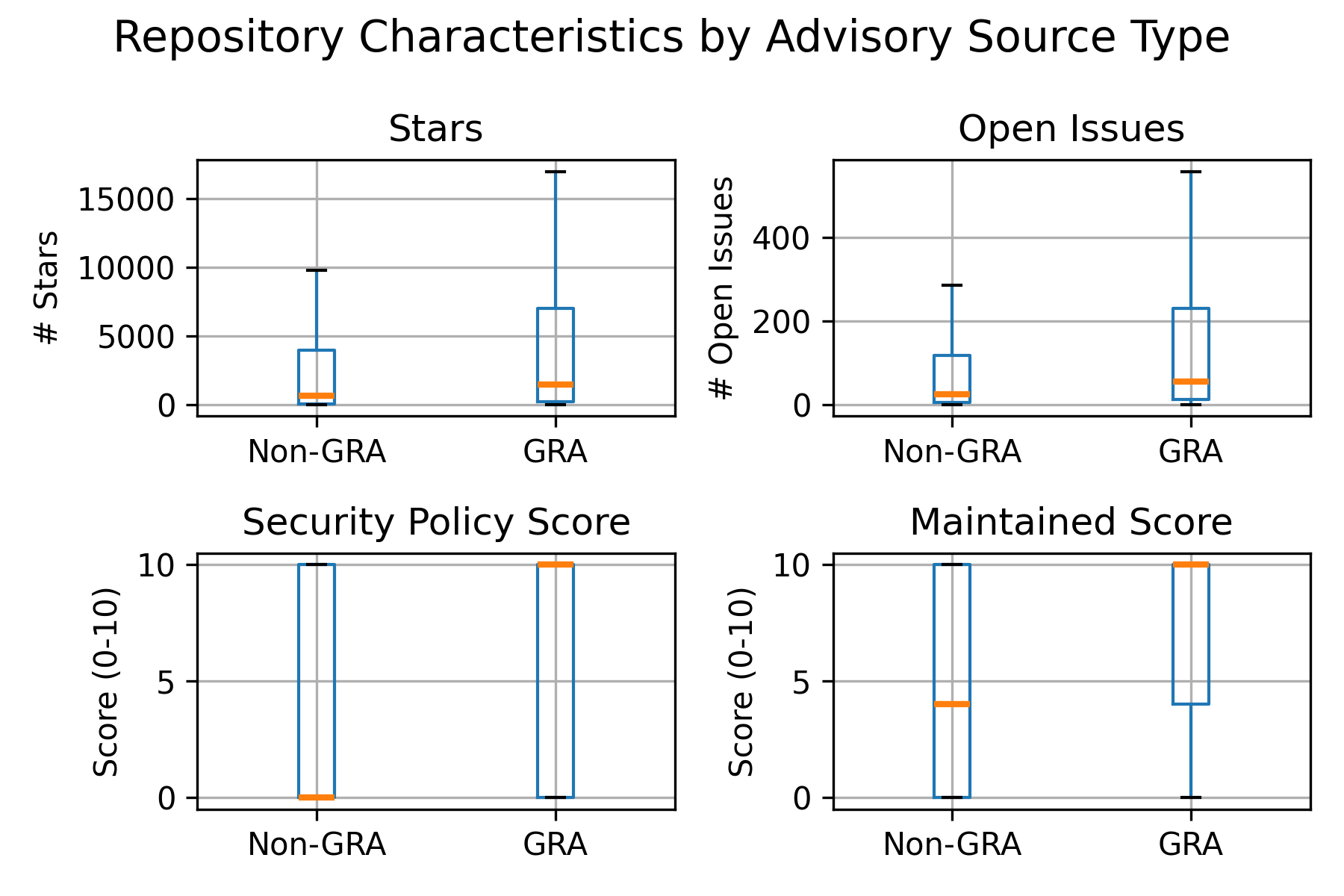}
    \caption{Repository characteristics box plot.}    \label{fig:repo_characteristics_boxplot}
\end{figure}

\subsubsection{Analysis} Figure~\ref{fig:repo_characteristics_boxplot} compares GRA-linked and non-GRA-linked repositories. Across all metrics, GRA-linked repositories exhibit higher medians. Although variability is large in all dimensions, Mann–Whitney tests confirm that the differences are statistically significant ($p < 0.0001$) with small-to-moderate effect sizes ($|\mathrm{RBC}|$ up to 0.3). In particular, GRA-linked repositories are more likely to maintain explicit security policies ($\mathrm{RBC} = -0.3$) and to be more actively maintained ($\mathrm{RBC} = -0.28$). Complete test statistics for these comparisons are provided in the replication package.

We also found that, among the repositories for which we collected the Security-Policy score, 69.3\% of GRA-linked repositories have an explicit security policy (i.e., a Security-Policy score greater than zero)\cite{OpenSSFScorecard}, compared to 39.5\% among non-GRA repositories.

\begin{keyfinding}{Key finding 1 for RQ4}
Nearly 70\% of repositories associated with GRAs include an explicit security policy, compared to only about 40\% among non-GRA repositories.
\end{keyfinding}

\begin{figure}
    \centering
    \includegraphics[width=\linewidth]{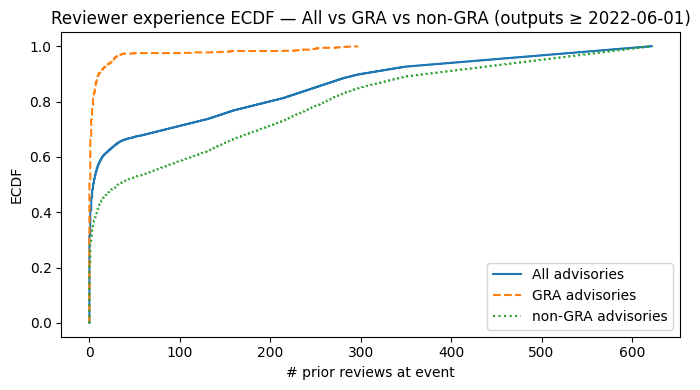} \vspace{-0.1in}
    \caption{Reviewer experience at review time (ECDF) for GRA vs.\ non-GRA advisories, and for all advisories combined.}
    \label{fig:reviewer_experience_ecdf} \vspace{-0.2in}
\end{figure}

Furthermore, to search for additional repository differences, we examined the \emph{reviewer experience at review time} for GRAs versus other advisories. For each credited reviewer–advisory event, we counted how many GHSA reviews that reviewer had completed \emph{before} that event (experience). The resulting empirical cumulative distribution functions (ECDFs) are shown in Figure~\ref{fig:reviewer_experience_ecdf}. To ensure consistency with the stabilized advisory pipeline following GitHub’s automation and NVD backfill, we restrict outputs to reviews after June 2022, while experience counts include all prior history.

The distributions show a notable difference between the groups. For GRA advisories the median prior experience is \textbf{0} (at least half of GRA reviews are a reviewer’s first-ever credited GHSA review), and the 90th percentile is \textbf{11}. In contrast, non-GRA advisories have a median of \textbf{33} prior reviews and a 90th percentile of \textbf{379}. Means are likewise separated (GRA: \textbf{8.3} vs.\ non-GRA: \textbf{130}) with higher maxima among non-GRA (297 vs.\ 622). Visually, the GRA ECDF rises sharply and saturates early, indicating that GRAs are predominantly handled by first-time or low-experience reviewers—consistent with the notion that GRAs are initiated and resolved by repository maintainers. The non-GRA ECDF increases more gradually, reflecting a heavy tail of experienced reviewers.

\begin{keyfinding}{Key finding 2 for RQ4}
GRAs are mostly reviewed by low-experience contributors (median prior reviews = 0).
\end{keyfinding}

\section{RQ5--RQ7: Empirical Characterization of Review Pipeline} \label{sec:empirical}\label{sec:RQ5_RQ7}

Understanding how advisories propagate through the GHSA review process requires a clear empirical picture of their distribution, sources, and timing. In this section, we characterize the advisory review pipeline at scale, examining which advisories are reviewed, where they come from, and how their flows evolve over time. This provides the foundation for identifying fast and slow review paths and for modeling the underlying dynamics in later sections.

\subsection{Overview}

Table~\ref{tab:ecosystems} provides an overview of both the \textit{reviewed} and the \textit{unreviewed} advisories, broken down by ecosystem, severity and source. Our dataset spans the period from 2019 to 2025, and includes
a total of 288{,}604 advisories. Out of these, 23,563 advisories
(8.2\%) were reviewed by GitHub, while 265{,}041
(91.8\%) remain unreviewed.

\begin{table}[t]
\caption{Ecosystem of GitHub Security Advisories: reviewed advisories (except for last row), categorized by source and CVSS severity (C = Critical, H = High, M = Medium, L = Low).}
\label{tab:ecosystems}
\footnotesize
\begingroup
\setlength{\aboverulesep}{0pt}
\setlength{\belowrulesep}{0pt}
\setlength{\tabcolsep}{3.5pt} 
\rowcolors{3}{gray!10}{white}
\begin{tabular}{lrrrrrrrr}
\toprule
\textbf{Ecosystem} & \textbf{Total} & \multicolumn{3}{c}{\textbf{Source}}    &   \multicolumn{4}{c}{\textbf{Severity}} \\
\cmidrule(lr){3-5} \cmidrule(lr){6-9}
 &  &  \textbf{GRA} & \textbf{NVD} & \textbf{Other}  & \textbf{C} & \textbf{H} & \textbf{M} & \textbf{L} \\
\midrule
Maven      & 5866 & 781  & 5074  & 11  & 796  & 1924 & 2828 & 318 \\
Composer   & 4838 & 1061 & 3355  & 422 & 514  & 1283 & 2761 & 280 \\
Npm        & 4082 & 1080 & 2220  & 782 & 1010 & 1578 & 1284 & 210 \\
Pip        & 3872 & 1378 & 2465  & 29  & 497  & 1490 & 1578 & 307 \\
Go         & 2454 & 1181 & 1262  & 11  & 260  & 885  & 1119 & 190 \\
Rust       & 1010 & 313  & 475   & 222 & 159  & 377  & 369  & 105 \\
Rubygems   & 943  & 243  & 659   & 41  & 116  & 312  & 448  & 67 \\
Nuget      & 723  & 214  & 504   & 5   & 49   & 436  & 211  & 27 \\
Swift      & 39   & 23   & 16    & 0   & 2    & 21   & 14   & 2 \\
Erlang     & 36   & 10   & 24    & 2   & 5    & 10   & 16   & 5 \\
Actions    & 33   & 31   & 2     & 0   & 7    & 17   & 7    & 2 \\
Pub        & 12   & 5    & 7     & 0   & 0    & 6    & 3    & 3 \\
\midrule
Reviewed   & 23563  & 6220 & 15825 & 1518 & 3391 & 8243 & 10418 & 1511 \\
Unreviewed & 265041 & 0    & 265041& 0    & 23129& 102005& 124700& 10775 \\
\bottomrule
\end{tabular}
\endgroup
\end{table}

Concerning ecosystems, among the reviewed advisories, the Maven, composer and npm ecosystems dominate the dataset.
Regarding severities, as the table shows, the majority of advisories are classified in the GHSA database as Medium (M) or High (H), with Critical (C) and Low (L) severity advisories representing a smaller but non-negligible fraction. These severity levels correspond to the qualitative severity rating scale defined in the Common Vulnerability Scoring System (CVSS)~\cite{AboutGithubAdvisoryDatabase}.

We also distinguish advisories by   source. Among the reviewed advisories,
6{,}220 advisories (26.4\%) originated as GRAs, while 17{,}343 (73.6\%)
did not. In addition, 15{,}825 reviewed
advisories (67.2\%) originated from the NVD.

For the remainder of this work, we focus on \textbf{\textit{reviewed}} advisories. The following subsections address our research questions about their timing and propagation dynamics (RQ5--RQ7).

\subsection{Sources of Advisories}

\textbf{RQ5: } \emph{ What are the main sources of GitHub-reviewed advisories, and what is their relative contribution to the review process?}

We examine the sources of reviewed advisories and their relative contribution to the review process. The platforms that we consider in this analysis are: GHSA, GRA, NVD, FriendsOfPHP, RustSec, PyPA, RubySec and GoVulnDB, with NVD and GRA being the two most prevalent ones \cite{GHSADByTheNumbers}.
This list constitutes all external databases listed by GitHub as sources of GHSAs~\cite{AboutGithubAdvisoryDatabase}.

\subsubsection{Longitudinal Analysis}

\begin{figure}[!htbp]
  \centering
  \includegraphics[width=\linewidth]{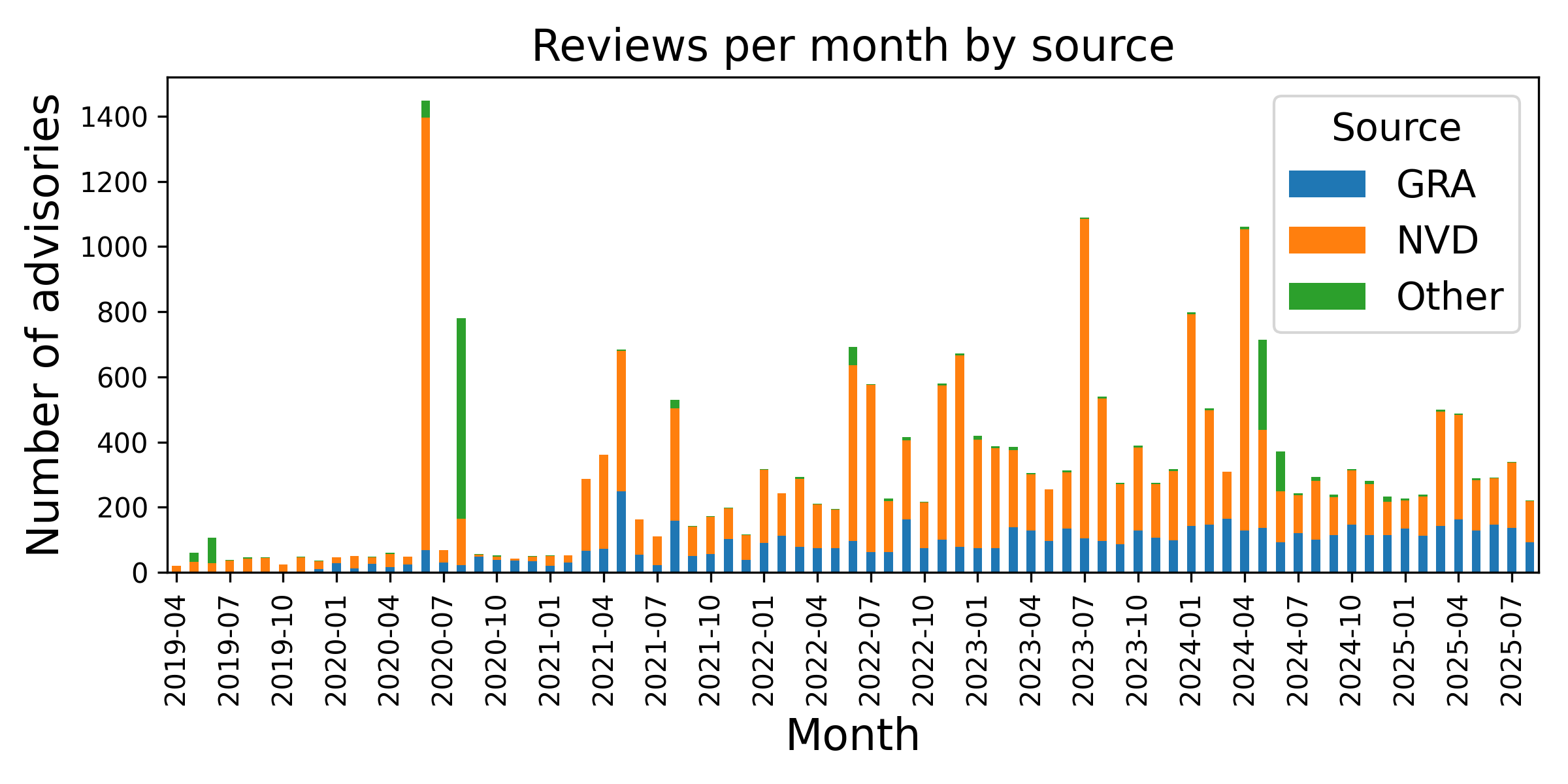}
  \caption{Reviews per month by source}
  \label{fig:reviews-per-month}
\end{figure}

Figure~\ref{fig:reviews-per-month} reports a longitudinal analysis of the advisories. Along with Table~\ref{tab:ecosystems}, it illustrates that GRAs are both relevant and stable over time—a significant portion of GHSAs originate from GRAs.  
Yet, despite their prevalence, there is almost no literature dedicated to GRAs.

\begin{keyfinding}{Key finding 1 for RQ5}
GRAs constitute a stable and persistent
source of GHSAs, accounting for roughly one-quarter all reviewed advisories.
\end{keyfinding}

\subsubsection{Flow of Advisories} \label{sec:flow}

\begin{figure}[h!]
  \centering
  \includegraphics[width=\linewidth]{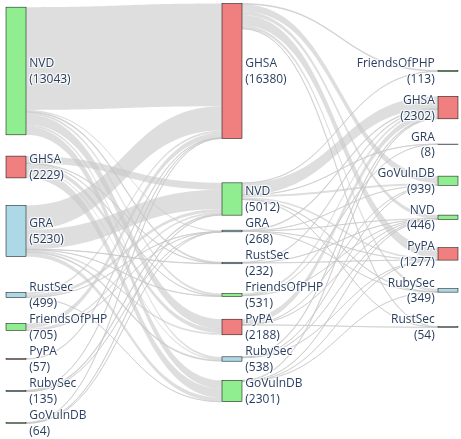}
  \caption{Flow of reviewed GHSA across platforms.}
  \label{fig:sankey}
\end{figure}

The advisory flow, i.e. the sequence in which advisories appear across platforms, can be represented as a Sankey diagram (Fig.~\ref{fig:sankey}). We use a methodology similar to that described by Miranda \textit{et al.}~\cite{OnTheFlowOfSecurityAdvisories}.

For each advisory, we consider the sequence of platforms that published it, sorted by publish date. From the sequence, we determine tuples $(x, y)$ where $x$ and $y$ are platforms with distinct consecutive publish dates. The Sankey diagram is generated from the multiset $\mathcal{S}$ of all such tuples, constructed as follows. If platforms $x$, $y$ and $z$ publish the same advisory on three distinct dates, we add tuples $(\mathrm{x}, \mathrm{y})$ and $(\mathrm{y}, \mathrm{z})$ to $\mathcal{S}$. If two platforms $x$ and $y$ publish on the same date, they are treated as simultaneous and no tuple is added for that pair. However, if a third platform $z$ publishes later, we add $(x, z)$ and $(y, z)$ to $\mathcal{S}$. Similarly, if a platform $w$ publishes before $x$ and $y$, we add $(w, x)$ and $(w, y)$.

The width of each link in Figure~\ref{fig:sankey} is proportional to the frequency of the tuples, and represents the intensity of advisory flow between the platforms. The diagram is organized into three levels: links from level~1 to~2 represent initial publication flows, while those from level~2 to~3 represent subsequent propagation steps.

We consider only advisories published both in GHSA and in at least one other platform on a distinct date.

\begin{keyfinding}{Key finding 2 for RQ5}
Approximately 95\% ($5230/(5230+268+8)$) of all advisory flows that traverse the GRA platform originate from it, following consistent propagation paths from GRA~$\rightarrow$~GHSA and from GRA~$\rightarrow$~NVD~$\rightarrow$~GHSA.
\end{keyfinding}

\subsection{Time from Publication to Review} \label{sec:timetopubtorev}

\textbf{RQ6: } \emph{What factors influence the time to review an advisory once it is published in GitHub?}

\subsubsection{Definition} We define \textbf{time to review} as the interval from an advisory’s publish date to its review date (\texttt{github\_reviewed\_at}). For advisories linked to a GRA, the publish date is taken from the GRA; otherwise, we use the GHSA \texttt{published\_at}.

\subsubsection{Data cleaning and filtering} In about 3{,}500 cases, the recorded review date precedes the publish date, resulting in negative review times. An inspection of these cases shows that they occur exclusively before April 2022, prior to the NVD backfill into GitHub \cite{GitHubBlogNVDBackfill}. Among these negative-time cases, approximately one third were published within one day after review, and a majority of those within a few hours, suggesting coordination effects or timestamp granularity issues (e.g., time zone differences) rather than substantive review-before-publication behavior. Since negative review times are artifacts of metadata timing and are not observed after the backfill cutoff, we restrict our analysis to advisories whose review date is not earlier than the publish date.

\begin{table}[!htbp]
\caption{Percentile distribution of time to review (measured in days) for GHSAs, by source.}
\label{tab:time_to_review_percentiles}
\footnotesize
\begin{tabular}{lrrrrrr}
\toprule
\textbf{Source} & \textbf{P25} & \textbf{P50} & \textbf{P75} & \textbf{P90} & \textbf{P95} & \textbf{P99} \\
\midrule
GRA (all) & 0.22 & 0.64 &  2.07 & 7.09 & 40.33 &  747.58 \\
NVD (all) & 0.93 & 35.26 & 513.03 & 707.57 & 834.08 & 1074.36 \\
GRA post-June 2022 & 0.18 & 0.45 & 1.14 & 2.95 & 4.5 & 59.25 \\
NVD post-June 2022 & 0.21 & 0.84 & 3.86 & 13.55 & 146.32 & 637.98 \\
\midrule
All reviewed & 0.35 &   2.65 & 320.74 & 657.86 & 715.44 & 1068.59 \\
\bottomrule
\end{tabular}
\end{table}

\begin{table}[h!]
\caption{Mann--Whitney statistical tests for time to review.}
\label{tab:hypothesis_tests}
\footnotesize
\begin{tabular}{lrrrr}
\toprule
\textbf{Comparison} & \textbf{RBC} & $p$-value \\
\midrule
GRA (5357) vs NVD (13048) & 0.62 & $<0.0001$ \\
GRA (4350) vs NVD (6047) post-June 2022 & 0.202 & $<0.0001$ \\
\bottomrule
\end{tabular}
\end{table}

\subsubsection{Empirical distribution and statistical comparison} Table~\ref{tab:time_to_review_percentiles} reports time-to-review percentiles.

To assess whether the distributions differ across advisory sources, we test the null hypothesis $H_0$ that the review lag distributions for GRA and NVD advisories are identical. The alternative hypothesis $H_1$ is that they differ. Because review lag data are not normally distributed (D’Agostino’s $K^2$ test, $p < 0.0001$), we employ the non-parametric Mann–Whitney 
$U$ test, complemented by the rank-biserial correlation (RBC) to estimate effect size.

Table~\ref{tab:hypothesis_tests} reports the results for both the full dataset and the subset of advisories published after June 2022. Prior to testing, outliers were removed from each group (GRA and NVD)\footnote{Outliers are defined as values outside the range $[Q_1 - 1.5 \times \mathrm{IQR},; Q_3 + 1.5 \times \mathrm{IQR}]$, where $Q_1$ and $Q_3$ are the first and third quartiles, and $\mathrm{IQR} = Q_3 - Q_1$.}. The $U$ statistic is computed using the first sample listed (e.g., GRA vs.\ NVD) as the reference group. Hence, a positive RBC indicates shorter lags for that group.

The tests reject the null hypothesis ($p < 0.0001$). The positive $\mathrm{RBC}$\footnote{The $\mathrm{RBC}$ is computed using the formula from Wendt (1972) \cite{Wendt1972}, $r_U = 1 - 2U/(n_1 n_2)$, where we set $U$ as the Mann-Whitney statistic for the first sample.} indicates that the differences favor faster reviews for GRAs. In the post–June~2022 comparison, for instance, $\mathrm{RBC}=0.202$ implies that
$P(X_\mathrm{GRA} < X_\mathrm{NVD}) \;=\; \frac{\mathrm{RBC}+1}{2} \approx 60\%$, 
meaning that in nearly two out of three randomly chosen pairs, the GRA advisory is reviewed faster than the NVD advisory.

Additionally, for advisories published after June 2022, we found that, among GRAs, 4151 out of 4350 (95.4\%) were reviewed within five days. For NVD-linked advisories, 4731 out of 6047 (78.2\%) were reviewed that quickly.

\begin{keyfinding}{Key finding 1 for RQ6}
GRAs tend to be reviewed faster than NVD-imported advisories.
\end{keyfinding}

\subsubsection{Impact of NVD automation}

When analyzing the full dataset, as shown in Table~\ref{tab:time_to_review_percentiles}, GRAs are reviewed orders of magnitude faster than NVD advisories. The differences between GRA and NVD change noticeably for advisories published after June 2022, with a notable improvement for NVDs. This suggests that GitHub’s automation of NVD imports \cite{GitHubBlogUnreviewedAdvisories, GitHubBlogNVDBackfill} has significantly optimized the NVD advisory review process.

\begin{figure}[h]
  \centering
  \includegraphics[width=\linewidth]{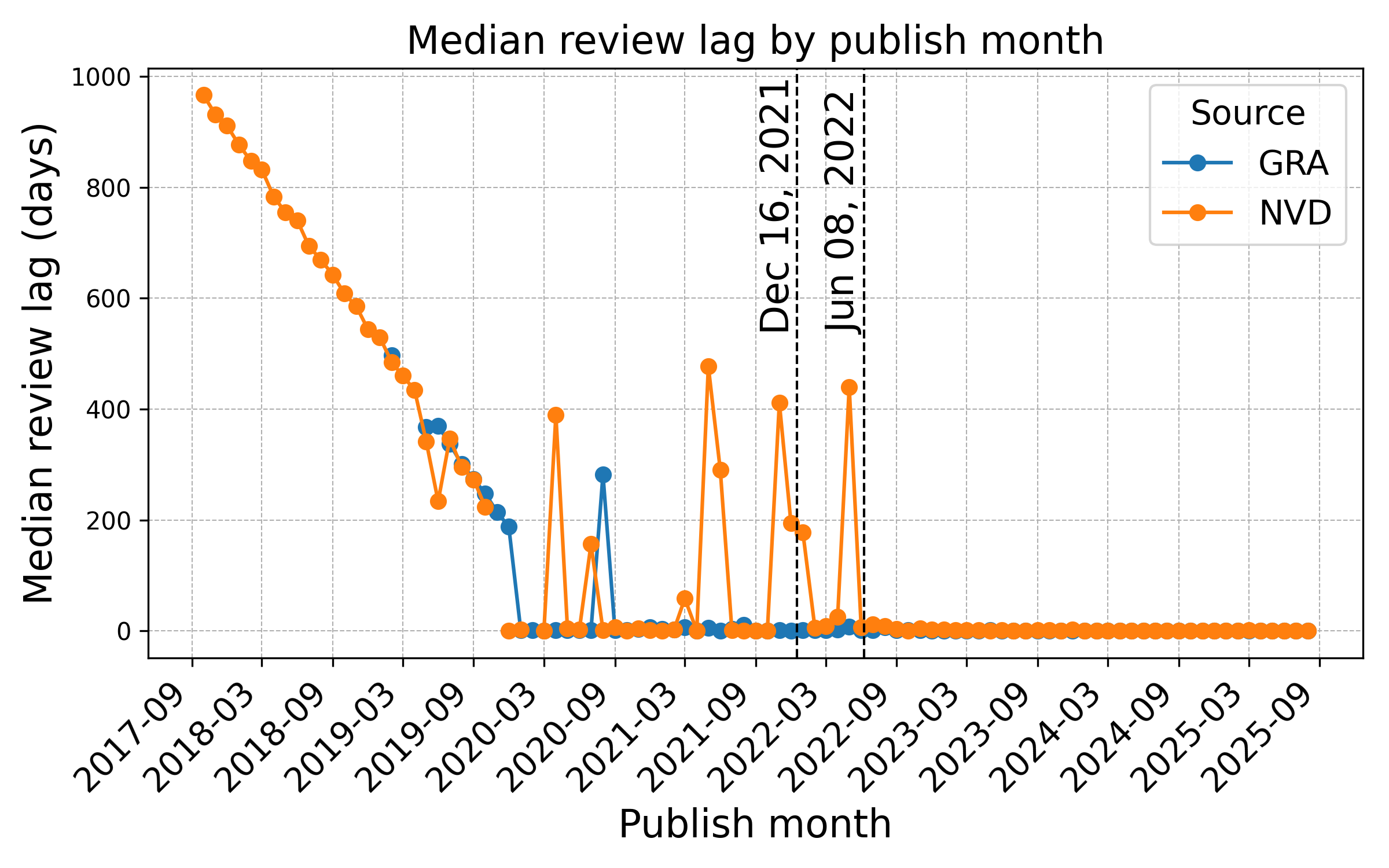}
  \caption{Median review time by publish month by source}
  \label{fig:lag_by_publish_month}
\end{figure}

Figure~\ref{fig:lag_by_publish_month} shows median review lags by publish month and by source. Although the automatic import process was announced in December 2021 \cite{GitHubBlogUnreviewedAdvisories}, a clear and sustained reduction in median review lags for NVD advisories is only observed after the NVD data backfill, which was announced in June 2022 \cite{GitHubBlogNVDBackfill}.
After this point, NVD advisories exhibit consistently short median lags, close to those of GRAs. This delayed effect is consistent with prior observations that infrastructure changes in software ecosystems may take time to propagate and stabilize \cite{Schorlemmer2024Signing}.

\begin{keyfinding}{Key finding 2 for RQ6}
The automation of the NVD-to-GHSA import process 
substantially altered the timing of advisory reviews.
\end{keyfinding}

\subsection{Time from Patch to Review}\label{sec:time_from_patch_to_review}

\textbf{RQ7: }
\emph{How long does it take for advisories to be reviewed after the corresponding patches are released?}

\subsubsection{Definition} In addition to review lag from publication, we analyze the time between a package’s first patched release and GitHub’s review of the advisory, which we call \textit{patch-to-review time}. Formally, this is $\texttt{github\_reviewed\_at} - \texttt{patched\_at}$, where \texttt{patched\_at} is the patch release date (see Section~\ref{sec:miningmethodology}). We obtained \texttt{patched\_at} for 10{,}532 advisories, from which we eliminated 453 (4.3\%) with a negative patch-to-review time.

\subsubsection{Why patch-to-review time matters?} From a defensive standpoint, the patch-to-review time reflects the period during which patch is available but likely not installed in a number of systems due to the lack of alerts~\cite{wang2017characterizing}. The fact that in GRAs the patch release seems to be coordinated with the publication of GHSAs suggests that the risks involved are significantly reduced compared to other sources such as NVD.  From an offensive standpoint, hackers can leverage patches to build exploits~\cite{ullah2025cve}. Therefore, the time between the patch release and the advisory review involves a race between the development of fully functional exploits, by hackers, and advisory reviews, by the security teams.

\begin{table}[h!]
\caption{Time from patch to review (days), after June 2022.}
\label{tab:patch_to_review_percentiles}
\footnotesize
\begin{tabular}{lrrrrrr}
\toprule
\textbf{Source} & \textbf{25th} & \textbf{50th} & \textbf{75th} & \textbf{90th} & \textbf{95th} & \textbf{99th} \\
\midrule
GRA & 0.51 &   2.04 &  12.18 &  78.99 &  129.62 &  582.14 \\
NVD & 7.06 &  27.69 & 127.23 & 672.7  & 1152.49 & 2257.88 \\
\midrule
All reviewed & 1.3  &   8.18 &  54.34 & 266.33 &  698.8  & 1786.79 \\
\bottomrule
\end{tabular}
\end{table}

\subsubsection{Influence of advisory source}
Table~\ref{tab:patch_to_review_percentiles} reports the time between patch release and GitHub review for advisories published after the NVD automation. GRAs are typically incorporated into the GHSA database soon after fixes become available, whereas NVD advisories exhibit substantially longer delays (median ~28 days). 

A Mann–Whitney $U$ test comparing 2,414 GRAs with 2,268 NVD-sourced advisories strongly rejects the null hypothesis of identical distributions ($p < 0.0001$), with $\mathrm{RBC}=0.559$, indicating that in roughly 78\% of randomly paired cases, the GRA patch-to-review time is shorter than its NVD counterpart. Full test statistics are available in the replication package.

\begin{keyfinding}{Key finding for RQ7}
GRAs are reviewed significantly sooner after patches than NVD-based advisories, with a median of 2~days versus 28~days and a strong effect size ($\mathrm{RBC}$=0.56). \end{keyfinding}

%
%

\section{Modeling Review Time} \label{sec:model}

 In this section, we develop a queueing-based abstraction of the GHSA review pipeline, grounded in the empirical   patterns observed in Section~\ref{sec:empirical}. This model captures how advisories flow from patch and publication events into the review queue, allowing us to explain fast and slow review paths and to support what-if analyses of disclosure strategies and reviewer capacity.


\begin{figure*}
    \centering
    \begin{tabular}{ccc}
    \includegraphics[width=0.3\linewidth]{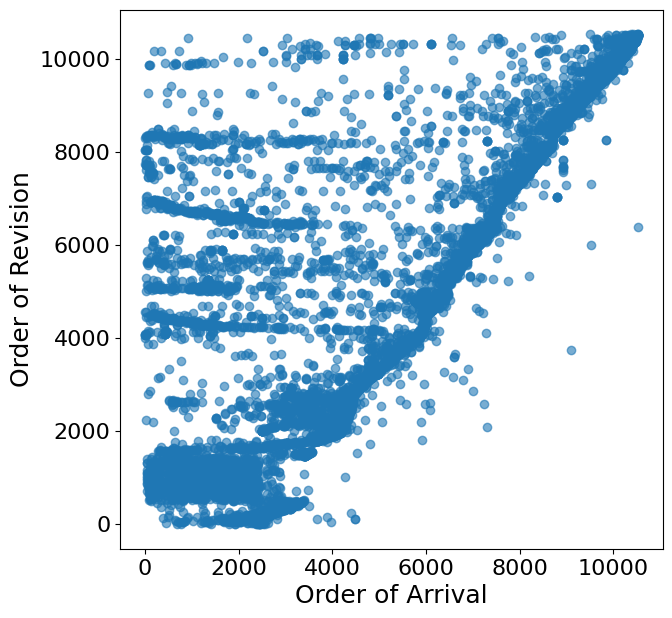} & \includegraphics[width=0.3\linewidth]{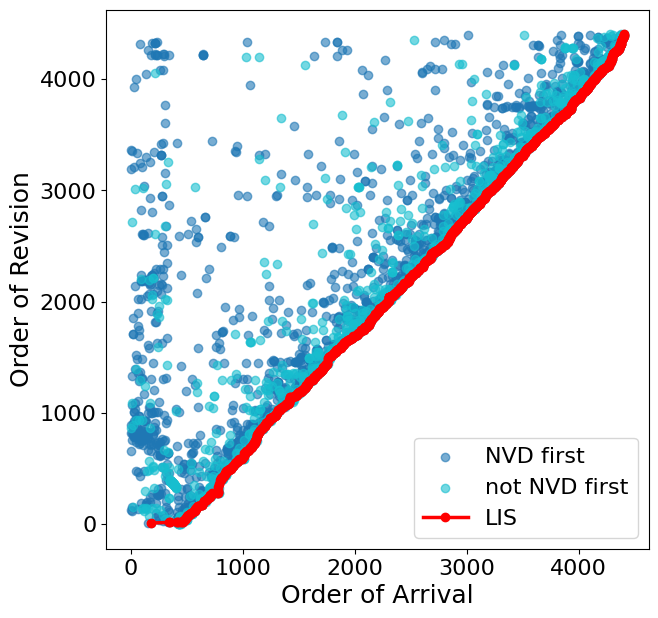} & \includegraphics[width=0.3\linewidth]{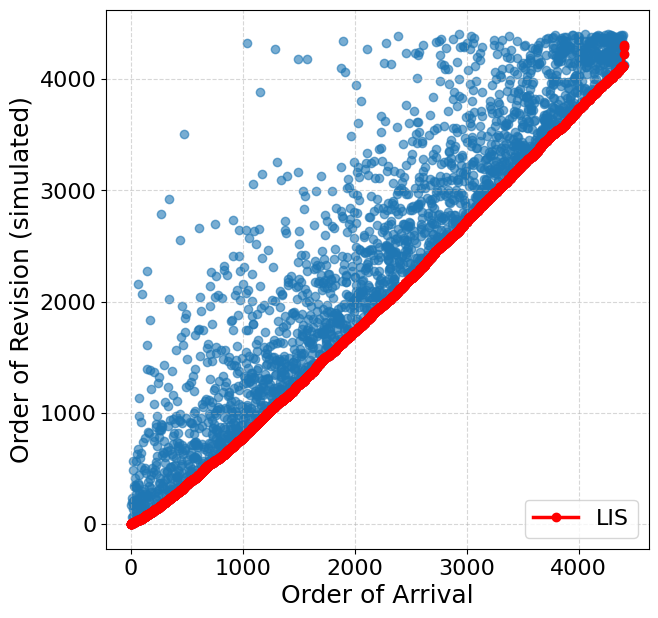} \\
    (a) & (b) & (c) 
    \end{tabular}  
    \caption{Advisory review  and publication order: (a) raw data (10,532 advisories); (b)   data after Jun 2022 (4,404); (c) model results.} 
    \label{fig:advisories}
\end{figure*}

\subsection{FIFO versus Random}\label{sec:6.1}

We begin by examining the relationship between the order of arrival and the order of review to assess whether the review process can be approximated by a simple FIFO queue model.

\subsubsection{Temporal ordering and impact of automation}
In this analysis, we assume that the life cycle of an advisory begins with the release of the patch addressing the reported vulnerability and ends with its review in the GHSA, i.e., the release of the patch corresponds to the advisory arrival into the system. Figure~\ref{fig:advisories}(a) presents the resulting scatter plot for the complete dataset, including advisories published before June 2022, while Figure~\ref{fig:advisories}(b) shows the scatter plot considering only advisories published from June 2022 onward. The more ordered behavior in Figure~\ref{fig:advisories}(b) compared to Figure~\ref{fig:advisories}(a) is consistent with the discussion in Section~\ref{sec:timetopubtorev}, where we highlight the impact of automation on review latency, and with the observation that the \emph{NVD-to-GHSA import process substantially altered the timing of advisory reviews}~\cite{GitHubBlogUnreviewedAdvisories,GitHubBlogNVDBackfill} (see key finding 2 for RQ6). For this reason, in the remainder of this section, we restrict our analysis to the dataset corresponding to Figure~\ref{fig:advisories}(b).

\subsubsection{FIFO evidence via Longest Increasing Subsequence (LIS)} \label{subsubsection:fifovslis}
To quantify how closely the review process follows FIFO, we compare the longest increasing subsequence (LIS) in the empirical review sequence, ordered by publication index ($x$-axis) and review index ($y$-axis), against the random-permutation baseline. For a uniformly random permutation of size $n$, $\mathbb{E}[L_n] \sim 2\sqrt{n}$ as $n \to \infty$~\cite{LoganShepp1977,VershikKerov1977}, so the LIS fraction $L_n/n$ scales as $2/\sqrt{n}$ and vanishes as $n$ grows. Our dataset exhibits an LIS of $\ell_{\text{obs}} = 984$ out of $n = 4404$ advisories (fraction $\approx 0.23$), substantially larger than the random baseline $2/\sqrt{n} \approx 0.03$, indicating strong FIFO tendency. A large LIS relative to baseline is consistent with predominantly FIFO discipline with minority exceptions. Indeed, from Figure ~\ref{fig:advisories}(b), we can see a clear diagonal line, demonstrating that a substantial fraction of  advisories were reviewed in the order in which their patches were released. 

\subsection{Queueing Model: Patch, Publish and Review}\label{sec:6.2}

\begin{figure}[!htbp]
    \centering
    \begin{tabular}{c}    \includegraphics[width=1\linewidth]{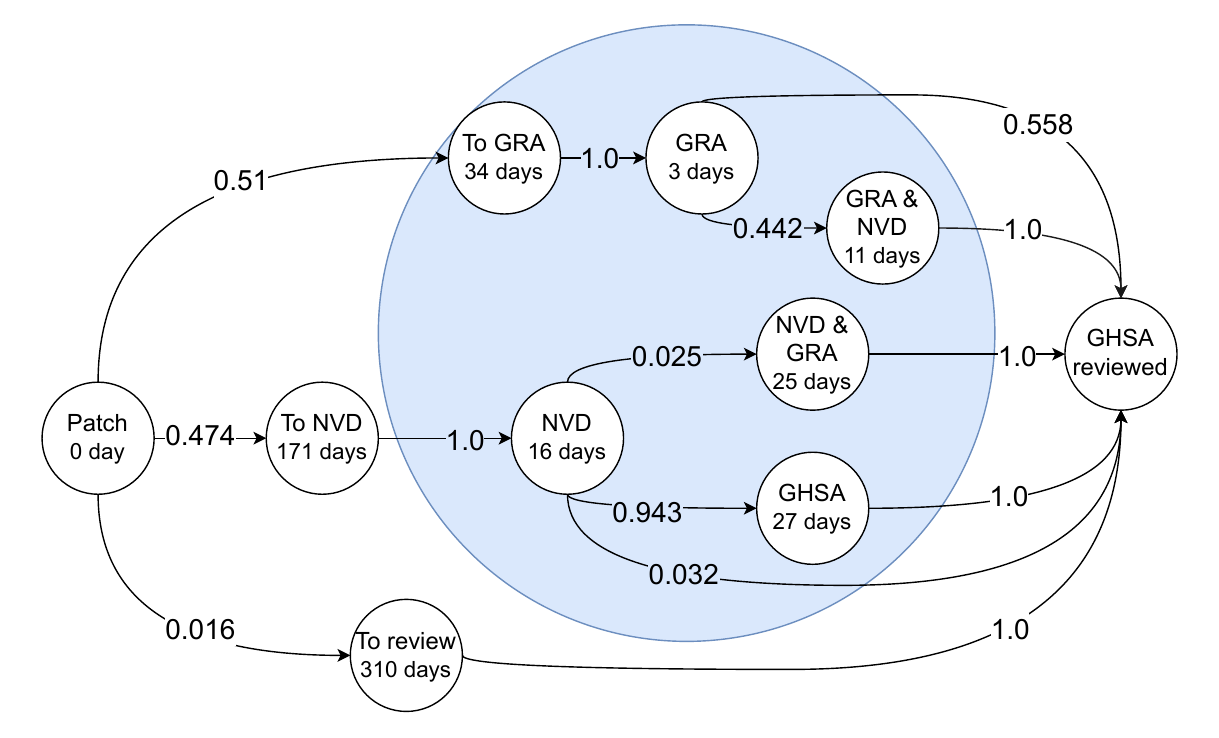}\\
    (a) \\    \includegraphics[width=\linewidth]{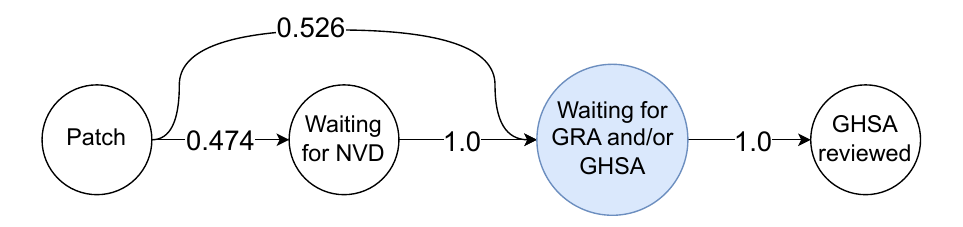}\\
    (b) \\   \includegraphics[width=\linewidth]{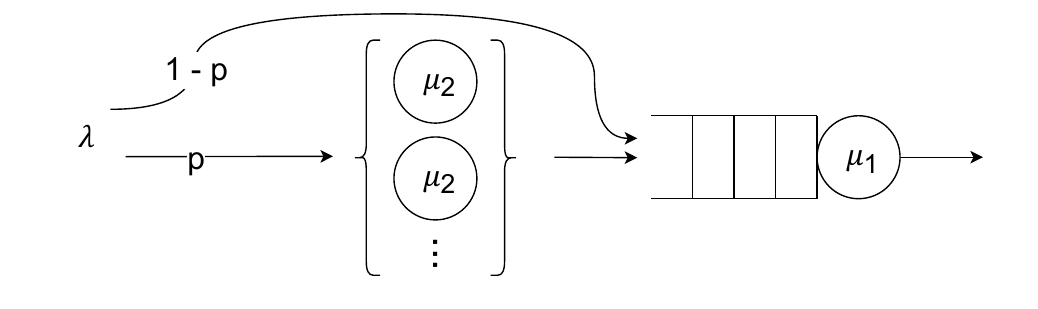} \\
    (c)\\
    \end{tabular} 
    \caption{Modeling the review process: (a) detailed and (b) simplified models, and (c)   corresponding queueing network.}  
    \label{fig:diagram}
\end{figure}

This section introduces a simple yet expressive queueing model of the advisory review process. The model captures key temporal patterns, quantifies the effect of disclosure paths on review latency, and supports what-if analyses for exploring potential policy or process changes.
To build this model, we consider the corresponding sequence of events occurring throughout an advisory's lifetime, including: (i) \emph{patch release date}, (ii) \emph{NVD publish date}, (iii) \emph{GRA publish date}, (iv) \emph{GHSA publish date}, and (v) \emph{review date}. These events capture the key steps that an advisory may follow from disclosure to   review.  From these sequences, we constructed a transition diagram that summarizes the advisory lifecycle across the dataset.  In Figure~\ref{fig:diagram}(a), each state represents one stage in the advisory lifecycle, and the edges are annotated with the corresponding state transition probabilities estimated from our dataset. The time values inside each node denote the average time advisories spend in that stage.




To make the dynamics more tractable, we grouped the states enclosed by the dashed circle in Figure~\ref{fig:diagram}(a), yielding a simplified four-state model shown in Figure~\ref{fig:diagram}(b). This abstraction highlights the two dominant pathways leading to review: (i) advisories that are published in the NVD before being reviewed, and (ii) advisories that are published through alternative channels (e.g., GRA or GHSA) and reviewed directly. This is in agreement with our key finding for RQ7, noting that Section~\ref{sec:time_from_patch_to_review} reports median values, whereas in this section we consider mean values.

The FIFO-like behavior observed in Section~\ref{sec:6.1}, combined with the transition structure in Fig.~\ref{fig:diagram}(b), motivates a stochastic queueing abstraction of the review process. Fig.~\ref{fig:diagram}(c) shows the corresponding queueing network: (i) an $M/M/\infty$ queue modeling the time to reach NVD followed by (ii) an $M/M/1$ queue modeling the   review stage.   
 Advisories published through GRA bypass the $M/M/\infty$ queue and are served directly in FIFO order. However, because the $M/M/\infty$ stage introduces random delays and reorderings, the overall process is no longer strictly FIFO. This aligns with the empirical patterns in Figure~\ref{fig:advisories}(b): while the LIS is large, indicating strong FIFO tendencies, a substantial fraction of advisories fall outside the LIS, reflecting the stochastic variability introduced by the NVD-first path.


\noindent\emph{\textbf{Model parameters.}} Advisories arrive as a Poisson process with rate $\lambda$. A fraction $p$ of them follow the NVD-first path, incurring an $M/M/\infty$ delay with service rate $\mu_2$ (mean $1/\mu_2$), while all advisories are ultimately reviewed through an $M/M/1$ FIFO queue.  Let $\mu_1$ be the service rate of such queue.

\noindent\emph{\textbf{Mean time to review.}}  
With probability $p$, advisories first experience the $M/M/\infty$ delay ($T_2 = 1/\mu_2$), then the $M/M/1$ delay ($T_1 = 1/(\mu_1 - \lambda)$)~\cite{harchol2013performance}. With probability $1 - p$, they go directly to the $M/M/1$. The expected review time is therefore
\begin{equation}
\label{eq:avg_review_time}
T = (1 - p)T_1 + p(T_2 + T_1) 
  = \frac{1}{\mu_1 - \lambda} + \frac{p}{\mu_2}.
\end{equation}

\noindent\emph{\textbf{Model parametrization.}} 
We estimated $\lambda$ from patch dates, excluding the 10th–90th percentile outliers, yielding $\lambda = 3.413$. The routing probability $p$ corresponds to the share of NVD-first advisories, $p = 0.474$. Parameters $\mu_1$ and $\mu_2$ were chosen to match observed mean review times: $49.79$ days (non-NVD-first) and $211.63$ days (NVD-first), giving $\mu_1 = 3.433$ and $\mu_2 = 0.006$.

Data generated by the model are shown in Figure~\ref{fig:advisories}(c). Comparing Figures~\ref{fig:advisories}(b) and~\ref{fig:advisories}(c), the model closely reproduces the observed trends. To validate this, we compared arrival–review rank differences between real and simulated data using a t-test; the $p$-value of $1.0$ indicates no significant difference. Other hypothesis tests yielded consistent results and are omitted due to space constraints.

\noindent\textbf{\emph{Fast and slow paths.}}
The model naturally explains the observed latency dichotomy. GRA-first advisories bypass the $M/M/\infty$ stage and enter directly into the $M/M/1$ queue, leading to short delays and FIFO-like behavior. NVD-first advisories, in contrast, experience an additional stochastic delay in the $M/M/\infty$ stage before review, producing two distinct latency regimes: a \emph{fast lane} (GRA-first) and a \emph{slow lane} (NVD-first). These patterns are an inherent consequence of the pipeline structure, not of explicit prioritization.

\noindent\textbf{\emph{What-if analysis.}}
Using Equation~\eqref{eq:avg_review_time}, we varied $p$ (the NVD-first fraction) while keeping other parameters fixed. Reducing $p$ from its baseline value of $0.474$ to $0.10$ (representing a scenario where 90\% of advisories originate as GRAs) lowers the average review time from $129$ to $66.7$ days, a reduction of nearly 50\%. This sensitivity analysis shows that the distribution of advisories across entry paths has a large effect on average waiting times in the modeled pipeline.

\section{Discussion and Implications} \label{sec:discussion}

\noindent\emph{--- \textbf{Fast and Slow Review Dynamics: GRAs against NVD Advisories.}}
The answers to RQ5, RQ6 and RQ7 reveal systematic differences between GitHub Repository Advisories (GRAs) and advisories sourced from the NVD in terms of review latency and its relationship to patch availability. After June 2022, GRAs exhibit a substantially shorter patch-to-review gap than NVD-sourced advisories, with median values of approximately 2 days versus 28 days. This indicates that, within the GRA workflow, advisory review tends to occur shortly after a fix becomes available.

\noindent\emph{--- \textbf{Implications for Automated Dependency Management Tools.}}
The observed alignment between patch availability and review timing has practical implications for tools that rely on reviewed GitHub Security Advisories (GHSAs). Dependabot consumes reviewed GHSAs to generate vulnerability alerts and remediation guidance. As a result, shorter patch-to-review gaps correspond to earlier availability of reviewed advisory metadata once a fix exists, which affects the timeliness with which automated dependency update information can be presented to developers.

\noindent\emph{--- \textbf{GHSAs as an Aggregation and Redistribution Hub.}}
Our Sankey flow analysis (Fig. ~\ref{fig:sankey}) highlights the role of the GHSA database as a central aggregation point for vulnerability information. It integrates advisories originating from multiple sources and serves as a redistribution hub within the vulnerability disclosure landscape. This structural position suggests that distinctions between reviewed and non-reviewed advisories at the GHSA level may shape how vulnerability information is exposed to downstream tools and services that reference GitHub's advisory ecosystem.

\noindent\emph{--- \textbf{Hypotheses on Timing Effects in the Vulnerability Ecosystem.}}
Taken together, the shorter review latency observed for GRAs and the hub role of GHSAs motivate hypotheses about broader timing effects in the vulnerability ecosystem. In particular, more tightly coupled review processes may support faster dissemination of patch-related vulnerability information across tools that consume GitHub advisories. Examining how these timing differences translate into dependency update behavior or patch adoption in practice remains an open direction for future work.

\noindent\emph{--- \textbf{Practical Implications for Advisory Review Processes.}}
From a practical perspective, our results emphasize the role of advisory review processes in shaping when vulnerability metadata becomes available to automated tooling. Review latency directly affects the temporal availability of actionable security information ~\cite{ayala2025deep,ayala2024poster,ayala2025investigating}, highlighting review efficiency as an important factor in the operation of modern software vulnerability ecosystems.

\section{Related Work}
\label{sec:relatedwork}

Security advisories play a key role in how vulnerabilities are disclosed, propagated, and reviewed. Prior work has studied different aspects of this process, including disclosure practices~\cite{arora2010empirical}, automation~\cite{he2023automating}, and governance~\cite{ayala2025mixed,alexopoulos2021vulnerability}.  Our work complements these efforts by examining the role of GHSA and GRAs,  and modeling how different disclosure paths affect review latency.

\subsection{Databases, Patching and Disclosure}

Research on vulnerability management has traditionally focused on public vulnerability databases such as the NVD, often examining severity scoring, disclosure timelines, and prioritization strategies~\cite{AlRubaye2020}. These studies emphasize how centralized vulnerability repositories shape patching behaviors and influence security response workflows, particularly through mechanisms such as CVSS scoring~\cite{allodi2020measuring} and vendor notifications~\cite{arora2010empirical}.

A foundational thread in this space investigates the characteristics of security patches and their role in the software supply chain. For example, Li and Paxson conducted one of the first large-scale empirical studies of security patches, highlighting how patching practices diverge from non-security changes in terms of complexity and latency~\cite{li2017large}. In parallel, Ponta et al. examined detection, assessment, and mitigation strategies for vulnerabilities in open-source dependencies~\cite{ponta2020detection}, while Decan et al. analyzed how vulnerabilities propagate through dependency networks, using the \texttt{npm} ecosystem as a prominent case study~\cite{decan2018impact} (see also~\cite{pinckney2023large}). These works collectively underscore the centrality of dependency relationships and patching workflows in modern vulnerability management.

Beyond patching behavior, a second line of work has investigated the vulnerability disclosure process itself. Liu et al. provided a large-scale empirical study of how open-source projects manage vulnerability disclosures, highlighting factors that influence disclosure latency and remediation~\cite{liu2025empirical}. Earlier, Gegick et al. demonstrated how text mining techniques can be leveraged to identify security-related bug reports in industrial settings~\cite{gegick2010identifying}, foreshadowing later approaches for advisory triage and classification.

More recently, attention has shifted toward the GHSA ecosystem, which plays an increasingly central role in open-source vulnerability disclosure and propagation~\cite{ayala2025deep,ayala2024poster,ayala2025investigating}. These works analyze aspects such as the relationship between advisories and CVEs, patterns of review activity, and metadata completeness. However, they largely focus on aggregated statistics and review outcomes rather than the advisory \emph{sources} or temporal \emph{flows} across platforms.

To the best of our knowledge, no previous work has systematically examined  GRAs, although they constitute the origin of a substantial fraction of GHSAs. Existing GHSA studies also tend to overlook (i) historical longitudinal analysis of review timelines and (ii) the role of advisory origin in shaping review latency. Our study fills this gap by analyzing how the source of first publication affects review speed, identifying two distinct paths in the review pipeline: fast (GRA-first) and slow (NVD-first).

 \subsection{Automation, Synchronization, and Latency}

How developers respond to automated security alerts and dependency update notifications? A number of studies~\cite{kinsman2021software,mirhosseini2017can,he2023automating}   highlight how automation affects patch adoption timelines, a factor closely tied to the GRA-first flows analyzed in our work.  Prior research has also explored consistency and timeliness across vulnerability databases~\cite{bozorgi2010beyond,allodi2017security}, providing context for the synchronization gaps we observe between NVD-first and GRA-first advisories.

Survival analysis has been used to model vulnerability and patching latency~\cite{nappa2015attack,przymus2023secret}, though no previous work  has applied these methods to advisory review pipelines. Such pipelines, in turn, impact the software supply chain security~\cite{zimmermann2019small}, where advisory timing directly relates to  perceived risk and financial implications~\cite{telang2007empirical}.  Our work complements such threads of research, by introducing an analytical model to characterize the review process.

Finally, complementary research has explored the social dimension of vulnerability handling, such as the roles and behaviors of contributors involved in reporting, triaging, and reviewing advisories~\cite{ayala2025mixed,alexopoulos2021vulnerability}. Governance and policy perspectives have also gained relevance, particularly as GHSA became a CNA under the CVE program~\cite{MITRECNAGuidance}, which formalizes disclosure responsibilities and coordination procedures. Our findings extend this line of work by linking user roles and advisory origins to structural patterns in the review process, offering a more complete view of how vulnerabilities propagate and are handled in practice.

\section{Threats to Validity}\label{sec:threatstovalidity}

Next, we discuss the main limitations our work along with the steps we took to mitigate their impact.

\noindent\emph{--- \textbf{External validity.}} Our findings are based on GHSA data as of August 2025, which may limit the generalizability to future iterations of the review process. To focus on the post-automation period, we restricted most analyses to advisories published after June 2022, which could reduce the representativeness of earlier advisories. However, we explicitly separated pre- and post-automation periods and observed that the fast/slow path dichotomy persists across both regimes, suggesting that the underlying patterns are robust. Similarly, our review-time analyses consider only reviewed advisories, introducing potential selection bias. Yet, the large volume of reviewed advisories and their wide distribution across ecosystems (Table~\ref{tab:ecosystems}) provide substantial coverage and support the validity of our conclusions. 

\noindent\emph{--- \textbf{Construct validity.}}
Some GHSA records contain missing or inconsistent fields, such as absent \texttt{nvd\_published\_at} timestamps, which could affect the precision of temporal analyses. We mitigated this issue by querying the NVD API to recover missing values where available ($\S$~\ref{subsec:DataEnrichment}). For the Sankey diagram (Fig.~\ref{fig:sankey}), publication dates were inferred from ecosystem-specific databases using commit timestamps. While these may not exactly match actual publication dates, they provide a conservative upper bound, implying that our estimates of propagation delays are more likely to be underestimates than overestimates.  Consequently, any remaining imprecision is unlikely to exaggerate our results. 

\noindent\emph{--- \textbf{Internal validity.}} 
Our patch-to-review analysis draws on the deps.dev API, which provides dependency data for six major ecosystems (PyPI, Go, RubyGems, npm, Maven, and NuGet). Since advisories from other ecosystems or unindexed packages are not included, the dataset may exhibit selection bias. Nonetheless, these six ecosystems collectively represent nearly half of all reviewed advisories (10,532 in total 44.7\%), offering broad and diverse coverage across the advisory landscape. For our repository characteristics analysis, metadata such as stars, forks, and OpenSSF scores are captured as point-in-time snapshots, which may not precisely reflect the repositories’ conditions when advisories were issued. However, this limitation is unlikely to affect our conclusions, as the analysis is cross-sectional (comparing GRA versus non-GRA repositories) rather than longitudinal. Consequently, any temporal measurement bias applies uniformly across both groups and does not undermine the validity of the observed relative differences (Fig.~\ref{fig:repo_characteristics_boxplot}).

\noindent\emph{--- \textbf{Model validity.}} 
Our queueing model assumes that part of the review process follows a FIFO discipline   supported by empirical evidence (Section~\ref{subsubsection:fifovslis} and Fig.~\ref{fig:advisories}), but alternative prioritization mechanisms may coexist. While the model successfully reproduces the observed review-time distributions, it necessarily abstracts away certain operational factors such as batching, manual interventions, and queue saturation effects. Nevertheless, the LIS analysis (Section~\ref{sec:6.1}) provides quantitative support for predominant FIFO behavior, and the model’s ability to replicate empirical patterns and enable what-if analysis underscores its validity and practical usefulness despite these simplifications.

\section{Conclusion and Future Work} \label{sec:conclusion}

We presented a large-scale empirical study of the GHSA review pipeline, revealing two distinct review paths: \emph{fast} and \emph{slow}. GRAs follow a fast path, being reviewed earlier than NVD-first advisories, particularly from patch to review. A simple queueing model explains this behavior: GRAs bypass an initial buffering stage, while NVD-first advisories incur additional delays. This asymmetry in the pipeline structure produces systematically shorter review latencies for GRA-originating advisories, even in the absence of explicit prioritization.

Despite this speed differential, our information flow analysis shows that most advisories continue to follow other disclosure paths, indicating that the fast-review path remains underutilized. Future work includes investigating the organizational and technical barriers that shape disclosure-path selection, as well as exploring how coordinated disclosure policies and automation could reduce exposure windows across ecosystems. By combining empirical characterization with analytical modeling, this study contributes to a clearer understanding of how infrastructure design and disclosure pathways shape vulnerability review timelines in open-source ecosystems.

\section*{Acknowledgments}

This work was partially supported by INES.IA (National Institute of Science and Technology for Software Engineering Based on and for Artificial Intelligence), \url{www.ines.org.br}, CNPq grant 408817/2024-0. This work was also partially supported by CAPES and FAPERJ under grants E-26/204.268/2024 and E-26/260.168/2026, CNPq grants 444956/2024-7, 424622/2021-1 and 315106/2023-9, as well as Finep PlatCiber.


\bibliographystyle{ACM-Reference-Format}
\bibliography{main}

@inproceedings{AlRubaye2020,
  author    = {Abduljaleel Al-Rubaye and Gita Sukthankar},
  title     = {Scoring Popularity in GitHub},
  booktitle = {2020 International Conference on Computational Science and Computational Intelligence (CSCI)},
  year      = {2020},
  pages     = {222--227},
  publisher = {IEEE},
  address   = {Las Vegas, NV, USA},
  doi       = {10.1109/CSCI51800.2020.00044},
  isbn      = {978-1-7281-8707-0},
  url       = {https://doi.org/10.1109/CSCI51800.2020.00044}
}

@misc{GitHubAdvisoryDatabase,
  title        = {GitHub Advisory Database: Security vulnerability database inclusive of CVEs and GitHub-originated security advisories},
  author       = {{GitHub}},
  year         = {2025},
  url          = {https://github.com/advisories},
  note         = {See also \url{https://github.com/github/advisory-database} and \url{https://docs.github.com/en/code-security/security-advisories/working-with-repository-security-advisories/about-repository-security-advisories}. Accessed: 2025-08-18}
}

@inproceedings{li2017large,
  title={A large-scale empirical study of security patches},
  author={Li, Frank and Paxson, Vern},
  booktitle={Proceedings of the 2017 ACM SIGSAC Conference on Computer and Communications Security},
  pages={2201--2215},
  year={2017}
}

@article{ponta2020detection,
  title={Detection, assessment and mitigation of vulnerabilities in open source dependencies},
  author={Ponta, Serena Elisa and Plate, Henrik and Sabetta, Antonino},
  journal={Empirical Software Engineering},
  volume={25},
  number={5},
  pages={3175--3215},
  year={2020},
  publisher={Springer}
}

@article{liu2025empirical,
  title={An empirical study on vulnerability disclosure management of open source software systems},
  author={Liu, Shuhan and Zhou, Jiayuan and Hu, Xing and Cogo, Filipe Roseiro and Xia, Xin and Yang, Xiaohu},
  journal={ACM Transactions on Software Engineering and Methodology},
  year={2025},
  publisher={ACM New York, NY}
}

@inproceedings{decan2018impact,
  title={On the impact of security vulnerabilities in the npm package dependency network},
  author={Decan, Alexandre and Mens, Tom and Constantinou, Eleni},
  booktitle={Proceedings of the 15th international conference on mining software repositories},
  pages={181--191},
  year={2018}
}

@book{harchol2013performance,
  title={Performance modeling and design of computer systems: queueing theory in action},
  author={Harchol-Balter, Mor},
  year={2013},
  publisher={Cambridge University Press}
}

@inproceedings{pinckney2023large,
  title={A large scale analysis of semantic versioning in npm},
  author={Pinckney, Donald and Cassano, Federico and Guha, Arjun and Bell, Jonathan},
  booktitle={2023 IEEE/ACM 20th International Conference on Mining Software Repositories (MSR)},
  pages={485--497},
  year={2023},
  organization={IEEE}
}

@misc{MITRECNAGuidance,
  author       = {{MITRE Corporation}},
  title        = {CNA Rules and CVE Program Governance},
  year         = {2023},
  howpublished = {\url{https://www.cve.org/resourcessupport/allresources/cnarules}},
  note         = {Accessed: 2025-10-23}
}

@inproceedings{gegick2010identifying,
  title={Identifying security bug reports via text mining: An industrial case study},
  author={Gegick, Michael and Rotella, Pete and Xie, Tao},
  booktitle={2010 7th IEEE Working Conference on Mining Software Repositories (MSR 2010)},
  pages={11--20},
  year={2010},
  organization={IEEE}
}

@misc{OpenSSFScorecard,
  title        = {OpenSSF Scorecard Checks},
  author       = {{OSSF}},
  year         = {2025},
  url          = {https://github.com/ossf/scorecard/blob/main/docs/checks.md},
  note         = {Accessed: 2025-10-19}
}

@inproceedings{ayala2025mixed,
  title={{A Mixed-Methods  Study of Open-Source  Software Maintainers On Vulnerability Management and Platform Security Features}},
  author={Ayala, Jessy and Tung, Yu-Jye and Garcia, Joshua},
  booktitle={34th USENIX Security Symposium (USENIX Security 25)},
  pages={2105--2124},
  year={2025}
}

@inproceedings{ayala2025deep,
  title={A deep dive into how open-source project maintainers review and resolve bug bounty reports},
  author={Ayala, Jessy and Ngo, Steven and Garcia, Joshua},
  booktitle={2025 IEEE Symposium on Security and Privacy (SP)},
  pages={522--538},
  year={2025},
  organization={IEEE}
}

@inproceedings{alexopoulos2021vulnerability,
  title={{Who are vulnerability reporters? A large-scale empirical study on FLOSS}},
  author={Alexopoulos, Nikolaos and Meneely, Andrew and Arnouts, Dorian and M{\"u}hlh{\"a}user, Max},
  booktitle={Proceedings of the 15th ACM/IEEE international symposium on empirical software engineering and measurement (ESEM)},
  pages={1--12},
  year={2021}
}

@article{ayala2025investigating,
  title={Investigating Vulnerability Disclosures in Open-Source Software Using Bug Bounty Reports and Security Advisories},
  author={Ayala, Jessy and Tung, Yu-Jye and Garcia, Joshua},
  journal={arXiv preprint arXiv:2501.17748},
  year={2025}
}

@article{ullah2025cve,
  title={{From CVE Entries to Verifiable Exploits: An Automated Multi-Agent Framework for Reproducing CVEs}},
  author={Ullah, Saad and Balasubramanian, Praneeth and Guo, Wenbo and Burnett, Amanda and Pearce, Hammond and Kruegel, Christopher and Vigna, Giovanni and Stringhini, Gianluca},
  journal={arXiv preprint arXiv:2509.01835},
  year={2025}
}

@article{wang2017characterizing,
  title={Characterizing and modeling patching practices of industrial control systems},
  author={Wang, Brandon and Li, Xiaoye and de Aguiar, Leandro P and Menasche, Daniel S and Shafiq, Zubair},
  journal={Proceedings of the ACM on Measurement and Analysis of Computing Systems},
  volume={1},
  number={1},
  pages={1--23},
  year={2017},
  publisher={ACM New York, NY, USA}
}

@misc{GithubGlobalAdvisoriesAPI,
    title   = {REST API endpoints for global security advisories},
    author  = {{GitHub Docs}},
    year    = {2025},
    url     = {https://docs.github.com/en/rest/security-advisories/global-advisories?apiVersion=2022-11-28#list-global-security-advisories},
    note    = {Acessed: 2025-08-29}
}

@misc{AboutGithubAdvisoryDatabase,
    title  = {About the GitHub Advisory Database},
    author = {{GitHub Docs}},
    year   = {2025},
    url    = {https://docs.github.com/en/code-security/security-advisories/working-with-global-security-advisories-from-the-github-advisory-database/about-the-github-advisory-database},
    note   = {Accessed: 2025-10-16}
}

@inproceedings{bozorgi2010beyond,
  title={Beyond heuristics: learning to classify vulnerabilities and predict exploits},
  author={Bozorgi, Mehran and Saul, Lawrence K and Savage, Stefan and Voelker, Geoffrey M},
  booktitle={Proceedings of the 16th ACM SIGKDD international conference on Knowledge discovery and data mining},
  pages={105--114},
  year={2010}
}

@inproceedings{nappa2015attack,
  title={The attack of the clones: A study of the impact of shared code on vulnerability patching},
  author={Nappa, Antonio and Johnson, Richard and Bilge, Leyla and Caballero, Juan and Dumitras, Tudor},
  booktitle={2015 IEEE symposium on security and privacy},
  pages={692--708},
  year={2015},
  organization={IEEE}
}

@article{arora2010empirical,
  title={An empirical analysis of software vendors' patch release behavior: impact of vulnerability disclosure},
  author={Arora, Ashish and Krishnan, Ramayya and Telang, Rahul and Yang, Yubao},
  journal={Information Systems Research},
  volume={21},
  number={1},
  pages={115--132},
  year={2010},
  publisher={INFORMS}
}

@article{allodi2020measuring,
  title={Measuring the accuracy of software vulnerability assessments: experiments with students and professionals},
  author={Allodi, Luca and Cremonini, Marco and Massacci, Fabio and Shim, Woohyun},
  journal={Empirical Software Engineering},
  volume={25},
  number={2},
  pages={1063--1094},
  year={2020},
  publisher={Springer}
}

@inproceedings{zimmermann2019small,
  title={Small world with high risks: A study of security threats in the npm ecosystem},
  author={Zimmermann, Markus and Staicu, Cristian-Alexandru and Tenny, Cam and Pradel, Michael},
  booktitle={28th USENIX Security symposium (USENIX security 19)},
  pages={995--1010},
  year={2019}
}

@inproceedings{przymus2023secret,
  title={{The secret life of CVEs}},
  author={Przymus, Piotr and Fejzer, Miko{\l}aj and Nar{\k{e}}bski, Jakub and Stencel, Krzysztof},
  booktitle={2023 IEEE/ACM 20th International Conference on Mining Software Repositories (MSR)},
  pages={362--366},
  year={2023},
  organization={IEEE}
}

@article{telang2007empirical,
  title={An empirical analysis of the impact of software vulnerability announcements on firm stock price},
  author={Telang, Rahul and Wattal, Sunil},
  journal={IEEE Transactions on Software engineering},
  volume={33},
  number={8},
  pages={544--557},
  year={2007},
  publisher={IEEE}
}

@article{allodi2017security,
  title={Security events and vulnerability data for cybersecurity risk estimation},
  author={Allodi, Luca and Massacci, Fabio},
  journal={Risk Analysis},
  volume={37},
  number={8},
  pages={1606--1627},
  year={2017},
  publisher={Wiley Online Library}
}

@article{he2023automating,
  title={{Automating dependency updates in practice: An exploratory study on GitHub Dependabot}},
  author={He, Runzhi and He, Hao and Zhang, Yuxia and Zhou, Minghui},
  journal={IEEE Transactions on Software Engineering},
  volume={49},
  number={8},
  pages={4004--4022},
  year={2023},
  publisher={IEEE}
}

@inproceedings{mirhosseini2017can,
  title={Can automated pull requests encourage software developers to upgrade out-of-date dependencies?},
  author={Mirhosseini, Samim and Parnin, Chris},
  booktitle={2017 32nd IEEE/ACM international conference on automated software engineering (ASE)},
  pages={84--94},
  year={2017},
  organization={IEEE}
}

@inproceedings{kinsman2021software,
  title={How do software developers use github actions to automate their workflows?},
  author={Kinsman, Timothy and Wessel, Mairieli and Gerosa, Marco A and Treude, Christoph},
  booktitle={2021 IEEE/ACM 18th International Conference on Mining Software Repositories (MSR)},
  pages={420--431},
  year={2021},
  organization={IEEE}
}

@misc{AboutDependabotAlets,
    title  = {About Dependabot alerts},
    author = {{GitHub Docs}},
    year   = {2025},
    url    = {https://docs.github.com/en/code-security/dependabot/dependabot-alerts/about-dependabot-alerts},
    note   = {Accessed: 2025-10-16}
}

@misc{GithubBlogAdvisoryCredits,
    title = {Security advisories now have multiple types of credits},
    author = {{GitHub Blog}},
    year = {{2023}},
    url   = {https://github.blog/changelog/2023-03-07-security-advisories-now-have-multiple-types-of-credits/},
    note = {Accessed: 2025-10-16}
}

@misc{GithubDocsAboutCredits,
    title = {About credits for repository security advisories},
    author = {{GitHub Docs}},
    year = {{2025}},
    url   = {https://docs.github.com/en/code-security/security-advisories/working-with-repository-security-advisories/creating-a-repository-security-advisory#about-credits-for-repository-security-advisories},
    note = {Accessed: 2025-10-20}
}

@misc{GitHubBlogUnreviewedAdvisories,
  title        = {Advisory Database now includes an Unreviewed Advisories section},
  author       = {{GitHub Blog}},
  year         = {2021},
  url          = {https://github.blog/changelog/2021-12-16-advisory-database-now-includes-an-unreviewed-advisories-section/},
  note         = {Accessed: 2025-09-12}
}

@misc{GitHubBlogNVDBackfill,
  title        = {All historical NVD advisories are now listed on GitHub},
  author       = {{GitHub Blog}},
  year         = {2022},
  url          = {https://github.blog/changelog/2022-06-08-all-historical-nvd-advisories-are-now-listed-on-github/},
  note         = {Accessed: 2025-09-12}
}

@misc{GHSADByTheNumbers,
  title        = {GitHub Advisory Database by the numbers: Known security vulnerabilities and what you can do about them},
  author       = {Jonathan Evans},
  year         = {2025},
  url          = {https://github.blog/security/github-advisory-database-by-the-numbers-known-security-vulnerabilities-and-what-you-can-do-about-them/},
  note         = {Accessed: 2025-09-30}
}

@article{LoganShepp1977,
  author    = {Logan, B. F. and Shepp, L. A.},
  title     = {A variational problem for random Young tableaux},
  journal   = {Advances in Mathematics},
  volume    = {26},
  number    = {2},
  pages     = {206--222},
  year      = {1977},
  doi       = {10.1016/0001-8708(77)90030-5}
}

@article{VershikKerov1977,
  author    = {Vershik, A. M. and Kerov, S. V.},
  title     = {Asymptotics of the Plancherel measure of the symmetric group and the limit form of Young tableaux},
  journal   = {Soviet Mathematics Doklady},
  volume    = {18},
  pages     = {527--531},
  year      = {1977}
}

@inproceedings{kancharoendee2025categorizing,
  title={On Categorizing Open Source Software Security Vulnerability Reporting Mechanisms on GitHub},
  author={Kancharoendee, Sushawapak and Phichitphanphong, Thanat and Jongyingyos, Chanikarn and Reid, Brittany and Kula, Raula Gaikovina and Choetkiertikul, Morakot and Ragkhitwetsagul, Chaiyong and Sunetnanta, Thanwadee},
  booktitle={2025 IEEE International Conference on Software Analysis, Evolution and Reengineering (SANER)},
  pages={751--756},
  year={2025},
  organization={IEEE}
}

@inproceedings{ayala2024poster,
  title={{Poster: A glimpse of vulnerability disclosure behaviors and practices using GitHub projects}},
  author={Ayala, Jessy and Tung, Yu-Jye and Garcia, Joshua},
  booktitle={Proceedings of the 45th IEEE Symposium on Security and Privacy (S\&P)},
  year={2024}
}

@article{OnTheFlowOfSecurityAdvisories,
author = {Miranda, Lucas and Vieira, Daniel and Aguiar, Leandro and Menasché, Daniel and Bicudo, Miguel Angelo and Nogueira, Mateus and Martins, Matheus and Ventura, Leonardo and Senos, Lucas and Lovat, Enrico},
year = {2021},
month = {05},
pages = {1-1},
title = {On the Flow of Software Security Advisories},
volume = {PP},
journal = {IEEE Transactions on Network and Service Management},
doi = {10.1109/TNSM.2021.3078727}
}

@article{Wendt1972,
  author  = {Wendt, H. W.},
  title   = {Dealing with a common problem in Social Science: A simplified rank‐biserial coefficient of correlation based on the $U$ statistic},
  journal = {European Journal of Social Psychology},
  year    = {1972},
  volume  = {2},
  number  = {4},
  pages   = {463--465},
  doi     = {10.1002/ejsp.2420020412}
}

@article {cosentino2017,
  author       = { Valerio Cosentino and Javier L C{\'a}novas Izquierdo and Jordi Cabot },
  title        = { A systematic mapping study of software development with {GitHub} },
  journal      = { Ieee access },
  year         = 2017,
  pages        = { 7173--7192 },
  volume       = 5,
  publisher    = { IEEE }
}

@article {borges2018s,
  author       = { Hudson Borges and Marco Tulio Valente },
  title        = { What’s in a {GitHub} star? understanding repository starring practices in a social coding platform },
  journal      = { Journal of Systems and Software },
  year         = 2018,
  pages        = { 112--129 },
  volume       = 146,
  publisher    = { Elsevier }
}

@misc{NVD,
	author = {{NIST}},
	title = {{N}{V}{D} - {H}ome},
	howpublished = {\url{https://nvd.nist.gov/}},
	year = {2025},
	note = {[Accessed 23-10-2025]},
}

@inproceedings{Schorlemmer2024Signing,
  author       = {Taylor R. Schorlemmer and Kelechi G. Kalu and Luke Chigges and
                  Kyung Myung Ko and Eman Abu Ishgair and Saurabh Baghi and
                  Santiago Torres-Arias and James C. Davis},
  title        = {Signing in Four Public Software Package Registries: Quantity, Quality, and Influencing Factors},
  booktitle    = {Proceedings of the 45th IEEE Symposium on Security and Privacy (SP)},
  year         = {2024},
  pages        = {1160--1178},
  publisher    = {IEEE Computer Society},
  address      = {Los Alamitos, CA, USA},
  doi          = {10.1109/SP54263.2024.00215},
}

\end{document}